\documentclass[useAMS,usenatbib]{mn2e}

\usepackage{amsmath}
\usepackage{amssymb}
\usepackage{graphicx}
\usepackage{subfig}
\usepackage{float}

\def\Msol{\mathrel{\rm M_{\odot}}}

\def\um{\mathrel{\rm \mu m}}

\def\ergss{\mathrel{\rm erg \; s^{-1}}}
\def\lx{\mathrel{\rm L_{2-8keV}}}

\def\lir{\mathrel{\rm L_{IR,SF}}}
\def\liragn{\mathrel{\rm L_{IR,AGN}}}

\title[]{Deep ALMA photometry of distant X-ray AGN: improvements in star formation rate constraints, and AGN identification.}

\author[F. Stanley et al.]
    {\parbox[h]{\textwidth}{
        F. Stanley,$^{\! 1,2\, *}$
        C.~M. Harrison,$^{\! 3,2}$
        D.~M. Alexander,$^{\! 2}$
        J. Simpson,$^{\! 4}$
        K.~K. Knudsen,$^{\! 1}$
        J.~R. Mullaney,$^{\! 5}$
        D.~J. Rosario,$^{\! 2}$
        J. Scholtz$^{\! 2}$}
    \vspace*{6pt} \\
    $^1$Department of Space, Earth and Environment, Chalmers University of Technology, Onsala Space Observatory, SE-43992 Onsala, Sweden \\
    $^2$Centre for Extragalactic Astronomy, Department of Physics, Durham University, South Road, Durham, DH1 3LE, UK \\
    $^3$European Southern Observatory, Karl-Schwarzschild-Str. 2, 85748, Garching b. M{\"u}nchen, Germany \\
    $^4$Academia Sinica Institute of Astronomy and Astrophysics (ASIAA), No. 1, Section 4, Roosevelt Rd., Taipei 10617, Taiwan \\
    $^5$Department of Physics \& Astronomy, University of Sheffield, Hounsfield Road, Sheffield, S3 7RH, UK \\
    $^*$Email: flrstanley@gmail.com
    }

\begin{document}
\maketitle

\begin{abstract}
	We present the star formation rates (SFRs) of a sample of 109 galaxies with 
	X-ray selected active galactic nuclei (AGN) with moderate to high X-ray luminosities ($\lx  = 10^{42} - 10^{45} \ergss$),
	at redshifts $1 < z <4.7 $, that were selected to be faint or undetected in the {\it Herschel} bands. 
	We combine our deep ALMA continuum observations with deblended 8--500$\um$ photometry 
	from {\it Spitzer} and {\it Herschel}, and use infrared (IR) SED fitting and AGN -- star formation decomposition methods.  
	The addition of the ALMA photometry results in an order of magnitude more X-ray AGN in our sample with a measured 
	SFR (now 37\%). The remaining 63\% of the sources have SFR upper limits that are typically a factor of ~2-10 times 
	lower than the pre-ALMA constraints. With the improved constraints on the IR SEDs, 
	we can now identify a mid-IR (MIR) AGN component in 50\% of our sample, compared to only $\sim$1\% previously.
	 We further explore the F$_{870\um}$/F$_{24\um}$--redshift plane as 
	a tool for the identification of MIR emitting AGN, for three different samples representing AGN dominated, star formation dominated, 
	and composite sources. We demonstrate that the F$_{870\um}$/F$_{24\um}$--redshift plane can successfully split between AGN and star formation
	dominated sources, and can be used as an AGN identification method.  
	\end{abstract}

\begin{keywords}
 galaxies: star formation -- galaxies: active -- galaxies: evolution	
\end{keywords}

\section{Introduction}
The impact of the energetic output of a galaxy's active galactic nucleus (AGN)
on the surrounding interstellar medium (ISM), and the galaxy's ongoing star formation, 
is one of the main open questions in the study of galaxy evolution (e.g., see 
\citealt{Alexander12};  \citealt{Fabian12}; \citealt{Harrison17}, for reviews).
Studies of the mean star formation rates (SFRs) of distant X--ray AGN, have repeatedly
shown that, on average, AGN live in star-forming galaxies 
(e.g., \citealt{Lutz10};  \citealt{Shao10}; \citealt{Mullaney12a}; 
\citealt{Harrison12}; \citealt{Rosario12}; \citealt{Rosario13c}; \citealt{Azadi15}; 
\citealt{Stanley15}). Furthermore, studies looking into the trends of the mean 
SFRs as a function of X-ray and/or bolometric AGN luminosity appear to be, after some discrepancy 
(e.g., \citealt{Page12}; \citealt{Rosario12}; \citealt{Harrison12}), converging
to the conclusion that there is a flat trend between the mean SFRs as a function 
of AGN luminosity for X-ray selected AGN (e.g., \citealt{Azadi15}; 
\citealt{Stanley15}; \citealt{Lanzuisi17}). The flat trend has been interpreted as a result 
of the stochasticity of the AGN, that has shorter timescales than that of galaxy-wide SFR 
(e.g., \citealt{Gabor13}; \citealt{Hickox14}; \citealt{Volonteri15}; \citealt{Stanley15}; \citealt{Lanzuisi17}). 
Indeed, studies find a correlation when averaging the AGN luminosity, as a function of the SFR 
 \citep[e.g.][]{Rafferty11, Mullaney12b, Chen13, Delvecchio14}. This can be interpreted as 
evidence for an underlying longterm correlation of AGN activity and star formation (although see \citealt{McAlpine17} for an alternative explanation).   

Studies of luminous optical quasars have repeatedly found a positive trend between the mean SFRs as a function of bolometric AGN luminosity 
\citep[e.g.][]{Bonfield11,Rosario13c,Kalfountzou14,Gurkan15,Harris16,Stanley17}. 
However, in our recent study of \cite{Stanley17} we argue that the positive trend observed is not a
result of AGN-driven enhancement, but it is driven by galaxy properties such as stellar mass ($M_*$) and redshift ($z$) (also see \citealt{Yang17}).

The above observational evidence may lead to the conclusion that AGN have no effect on the SFR 
of their host galaxies. However, AGN feedback (i.e., where the large energy outputs of the AGN cause 
heating and/or outflows of a galaxy's gas) is a necessary component 
of cosmological hydrodynamical simulations of galaxy evolution 
(e.g., \citealt{Bower06}; \citealt{Genel14}; \citealt{Schaye15}). Recent work by 
\cite{McAlpine17} that studied the galaxies that host AGN in the EAGLE 
(i.e., Evolution and Assembly of GaLaxies and their Environments; \citealt{Schaye15}) simulation, 
that incorporates AGN feedback, successfully reproduces
the observational results of a flat trend between the mean SFR as a function of AGN luminosity 
for AGN selected samples, as well as those of a correlation of the mean AGN luminosity as a function of SFR. 
The fact that a simulation incorporating AGN feedback can reproduce the above flat trends 
demonstrates that we cannot rule out that AGN have an impact on their 
host galaxies (\citealt{Harrison17}). It may be that the signatures of AGN feedback are 
much more subtle than what is able to be traced by looking at the 
mean properties of AGN samples. However, the limitations presented by studying means, can be 
overcome by placing strong constraints on the underlying distribution (e.g., \citealt{Mullaney15}; Scholtz et al. 2018).

The main restriction in accurately measuring the distributions of SFRs of high 
redshift ($z > 0.2$) AGN samples, has been the limitations on the sensitivity of the available photometry. 
{\it Herschel} has provided the deepest field-survey observations in the far-infrared (FIR) at 
70--500$\um$, but even so the available 
surveys only detect the bright end of the galaxy population. For $z \gtrsim$1 we can only detect star-bursting and/or 
massive star-forming galaxies. Consequently, in order to directly constrain the SFRs of the typical population of galaxies 
and AGN at redshifts of $z \gtrsim 1$, we need even deeper observations in the FIR/sub-mm. Today, the Atacama large 
(sub-)mm array (ALMA) can achieve that. With ALMA it is now possible to easily detect and resolve galaxies at 
redshifts above $z\sim1$ at lower fluxes than that possible with previous FIR/Sub-mm observatories, and place more accurate 
constraints on the SFRs of fainter galaxies with and without AGN. This has 
been demonstrated  previously in \cite{Mullaney15}, where it was shown that with ALMA 
photometry it is possible to distinguish differences between the distribution of the 
SFRs of a sample of X-ray AGN, and that of the overall population of star-forming galaxies.
Despite the limited number of targets in the study of \cite{Mullaney15} (i.e., 24 X-ray AGN targets), the results highlight the importance 
of constraining the distribution of SFRs rather than just the mean. 

In this paper and the companion paper of Scholtz et al. (2018), 
we build on the sample of \cite{Mullaney15} with the observation 
of a larger sample of X-ray AGN covering higher X-ray hard-band (HB; 2-8keV) luminosities 
($\lx > 10^{44} \ergss$).
Here, we present a sample of 109 X-ray AGN observed with ALMA in Band-7 (i.e. 870$\um$), 
covering the redshifts of $1<z<4.7$ and X-ray HB luminosities of 
$10^{42} < \lx \leq 10^{45} \ergss$. 
An important factor that needs to be taken into account in such studies is 
the possible contribution of the AGN to the FIR/Sub-mm emission observed,
as argued by a number of AGN studies \citep[e.g.][]{Mullaney11,DelMoro13,Leipski13,Delvecchio14,Netzer16,Symeonidis16,Stanley17}. 
Especially when looking at FIR faint galaxies with deep ALMA observations, 
where AGN contamination could have a significant effect on the measured SFR values.
For this reason, we use available photometry covering 3.6--870$\um$, 
in order to perform individual SED fitting and decomposition of the 
star formation and AGN contributions to the IR SED, providing SFR constraints where
the AGN contamination has been removed as best as possible. 
The improved SFR measurements
presented here are used in the companion paper of Scholtz et al. (2018) 
in order to define the SFR and SFR/M$_*$ (sSFR) distributions of the AGN sample. 

In Section \ref{sample} we present the sample used for this study, 
and give information on the ALMA observations. In Section \ref{method}
we present the IR SED fitting method. In Section \ref{improvements}
we demonstrate the improvements on constraining the SFRs and identifying the AGN 
component of the IR SEDs that ALMA provides. In Section \ref{agn_comp} we demonstrate the use of the 
F$_{870\um}$/F$_{24\um}$--redshift plane as a selection tool for AGN. 
Finally, in Section \ref{conclusion} we give a summary of our results.
Throughout this paper we assume $\rm H_0 = 70 km \, s^{-1} \, Mpc^{-1}$,
$\rm \Omega_M = 0.3$, $\rm \Omega_\Lambda = 0.7$, and 
a \cite{Chabrier03} initial mass function (IMF).

\section{Sample \& Observations} \label{sample}
We present a sample of X-ray selected AGN that have been 
observed in two ALMA Band-7 programs during Cycle~1 and Cycle~2. 
Our ALMA Band-7 programs were designed with a key aim of constraining the 
SFR and sSFR distributions of a sample of X-ray AGN covering moderate to high 
X-ray luminosities, in the redshift range where we observe the peak of star formation and AGN 
activity. The details of the sample selection for the two programs are given in \cite{Scholtz18}, 
we give a brief overview in section~2.1.
Information on the ALMA observations and data reduction are given in section~2.2. The 
complementary MIR and FIR photometry for our sample is described in section~2.3. In 
section~2.4 we provide details on two comparison samples that are later used in section~5.

\subsection{Sample selection}
The Cycle~1 sample was selected from the 4Ms {\it Chandra} Deep Field South (CDF-S) catalogue of 
\cite{Xue11} to have $\lx > 10^{42} \ergss$ at redshifts of $1.5 < z < 3.2$ (see \citealt{Mullaney15}; \citealt{Harrison16}).
The sample was selected to be complete for host galaxy stellar masses of $>10^{10}M_\odot$.
The Cycle~2 sample was selected from the 1.8Ms {\it Chandra}-COSMOS (C-COSMOS) catalogues of \cite{Elvis09} and 
\cite{Civano12} covering the redshifts of $1.5 < z < 3.2$ and X-ray HB luminosities of $10^{43} < \lx \lesssim 10^{45} \ergss$, 
with a uniform sampling of the $\lx$--redshift plane in the above ranges. The luminosity range for this selection 
was chosen in order to cover the knee of the X-ray luminosity function at the redshifts of interest, i.e., L$* \sim 10^{44} \ergss$ at $z\sim2$ 
\citep[e.g.][]{Aird15}, and complement the Cycle~1 sample that covered
lower X-ray luminosities. The typical space densities of X-ray AGN at these luminosities 
and redshifts are $\sim$10$^{-4}$Mpc$^{-3}$ \citep[see Fig.~18 of][]{Aird15}. 

Both selections have been restricted to within the 
areas covered by the {\it Herschel} observational programs PEP/GOODS-\textit{H}
(\citealt{Lutz11}; \citealt{Elbaz11}) and HerMES (\citealt{Oliver12})
in the fields of GOODS-S, and COSMOS, which are our main sources of
the FIR photometry covering the observed wavelengths of 70 -- 500$\um$ (described in \citealt{Stanley15}). 
In both ALMA programs the targeted sources 
were primarily chosen to have insufficient {\it Herschel} photometry 
(i.e., detected in too few {\it Herschel} bands) to successfully constrain the IR SED and 
decompose it to the star-forming and AGN components. 
Consequently, our sample consists of mostly {\it Herschel}, 
and sometimes {\it Spitzer}, undetected sources with poor SFR constraints. 
We make use of the {\it Spitzer} and {\it Herschel} photometry assigned to the X-ray AGN 
in \cite{Stanley15} for our analysis (see section~2.3),
in combination with ALMA observations at 870$\um$. 
However, since the original selection of targets for our ALMA observed programs, 
new redshift catalogues of the 
CDF-S and C-COSMOS have been published by \cite{Hsu14}
and \cite{Marchesi16} respectively. 
In our analysis we make use of the updated redshifts from these catalogues.

In this paper we analyse the X-ray AGN that were observed by ALMA, 
including serendipitous detections within the ALMA primary beam, with $z >$ 1. 
This results in 109 X-ray AGN with ALMA 
870$\um$ observations, 101 originally targeted, and 8 serendipitous X-ray AGN.
There are an additional 5 sources with $z <$ 1 covered by the ALMA program, all in the field 
of GOODS-S, that are not included in the analysis of this paper, but their ALMA 
photometry and source properties are given in Scholtz et al. (2018).
Our sample covers an X-ray luminosity 
range of $10^{42} < \lx \lesssim 10^{45} \ergss$ and 
a redshift range of $1<z<4.7$.
In Fig.~1 we plot the $\lx$ as a function of redshift for 
the sample studied here, and highlight the ALMA 870$\um$ detected sources.
In Fig.~1 we also plot all X-ray AGN from the catalogues used 
in our selection in grey, as well as the L$_*$ track from \cite{Aird15}. 
It is easy to see that our sample covers almost the full luminosity range of 
the catalogued X-ray AGN at redshifts of $1.5<z<3.2$, and covers at least an order of magnitude 
on either side of the L$_*$, making it a representative sample of 
X-ray AGN at these redshifts. The luminosity range of our sample also 
covers the full range of X-ray luminosities typically 
included in studies of the SFR trends as a function of X-ray luminosity, 
and overlaps with the lower luminosities
of the more luminous quasars.

\begin{figure}
	\begin{center}
		\includegraphics[scale=0.5]{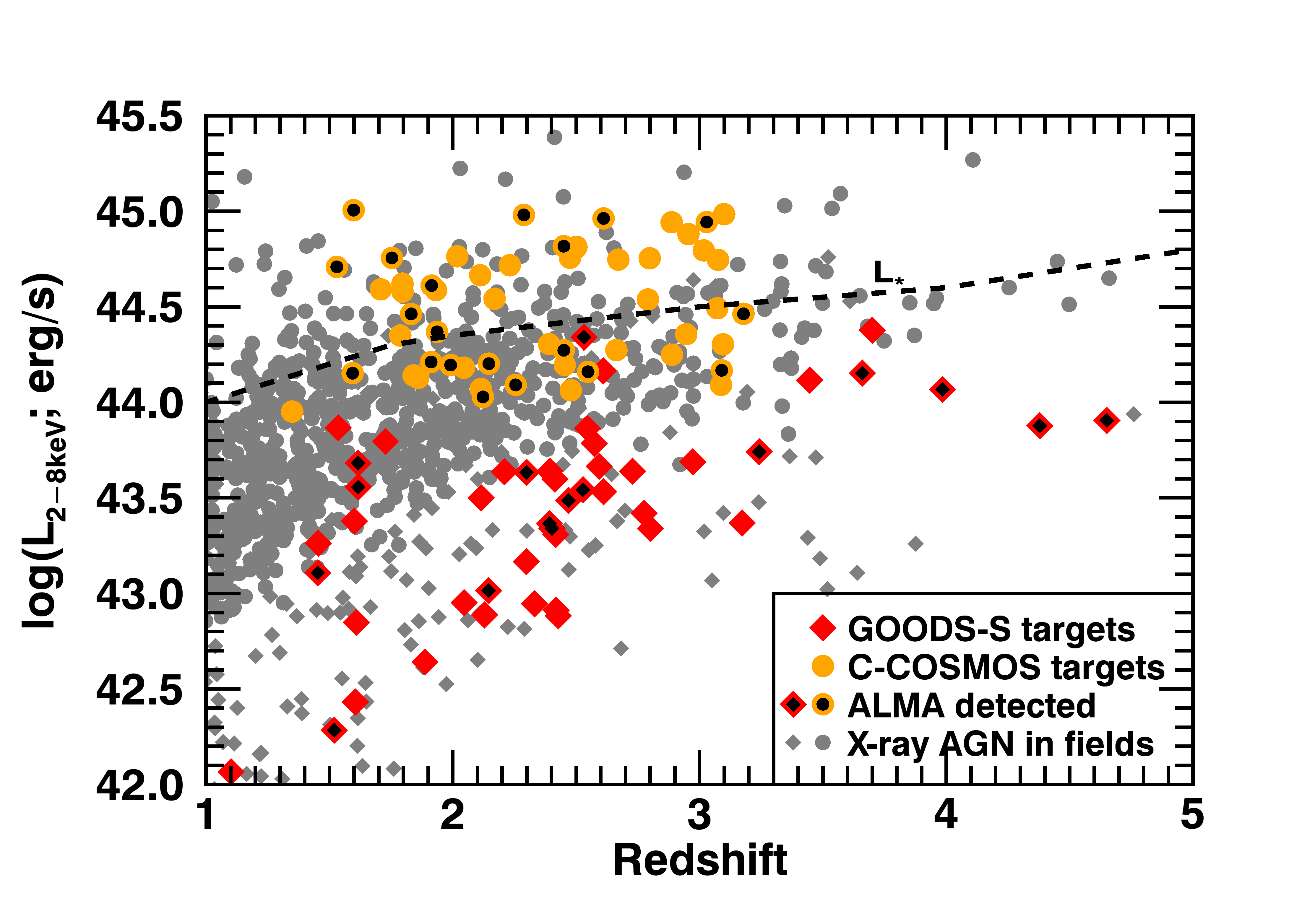}
		\caption{X-ray hard band (2--8keV) luminosity ($\lx$) as a function of redshift. 
			In grey we show all X-ray AGN catalogued in the GOODS-S (diamonds) and C-COSMOS (circles) 
			fields. In colour we show the 109 $z>1$ X-ray AGN observed with ALMA, including 101 originally targeted and 
			8 serendipitous detections. Detected sources are highlighted 
			with black centres. With the dashed curve we plot the knee of the X-ray luminosity function (L$*$) from \protect\cite{Aird15}.}
	\end{center}
\end{figure}

\subsection{ALMA 870um observations}
The sample of 109 X-ray AGN were observed during Cycle 1 (2012.1.00869.S; PI: J.~R.~Mullaney) 
and Cycle 2 (2013.1.00884.S; PI:D.~M.~Alexander) with a 
bandwidth of 7.5GHz centred at 351GHz, with 55 sources in CDF-S and 54 sources in C-COSMOS.
Cycle 1 observations were carried out using 32 antennas in the 12m array and 9 antennas in the 7m array, with 
integration times ranging between 2.5--13min. Cycle 2 observations were carried out using 34 antennas in the 
12m array and 9 antennas in the 7m array, with integration times ranging between 1--6minutes. 

The data were processed and imaged following the methods of \cite{Hodge13} and 
\cite{Simpson15}; see full details in \cite{Scholtz18}.
We used the {\sc common astronomy software application} 
({\sc casa}; version 4.4.0; McMullin et al 2007), and the {\sc clean} routine in {\sc casa}.
The raw data was calibrated using the ALMA data reduction pipeline. 
The results were visually inspected, and when deemed necessary,
the pipeline calibration process was repeated with additional data flagging.  
We created ``dirty" images, which were subsequently cleaned down to 3$\sigma$. 
We then identified the sources with SNR$\geq$5. To ensure the recovering of extended flux, 
we applied natural weighting 
and appropriate Gaussian tapering in the {\it uv}--plane to obtain a synthesised beam of $\sim$0.8$"$ for all images. 
The resulting synthesised beams are of the size of $(0.8"-0.9")\times0.7"$, 
with noise levels of 0.1--0.8 mJy/beam in CDF-S, 
and 0.08--0.23 mJy/beam in C-COSMOS. 
The large noise levels of 0.8mJy/beam correspond
to a sub-sample of 14 targets in the CDF-S field that were observed at higher resolution than that requested (i.e, 0.3" 
instead of 1" resolution). Therefore, for these observations the images had to be heavily tapered 
to a resolution of 0.8$"$, resulting in increased noise levels (see Scholtz et al. 2018). This is taken 
into account in section~4, when assessing the improvements in SED fitting due to ALMA photometry.

In Scholtz et al. (2018) we present the ALMA 870$\um$ photometric 
catalogues for the full sample of targeted and serendipitous X-ray AGN, along with a
detailed description of the catalogue production. 
The catalogue includes all targeted sources and serendipitous detections. 
If a source remains undetected we take 3$\times$RMS as the flux density upper limit.
In total we find that 40/109 (36.7\%) of our sources are detected by ALMA. 

\subsection{MIR and FIR photometry}

For our SED fitting analysis, we exploit available photometry in the 
observed frame wavelength range of 3.6 -- 500$\um$,
provided by observations carried out by: \textit{Spitzer}-IRAC at
3.6--8$\um$; \textit{Spitzer}-IRS at 16$\um$; \textit{Spitzer}-MIPS at
24$\um$; \textit{Herschel}-PACS at 70, 100, 160$\um$; and 
\textit{Herschel}-SPIRE at 250, 350, 500$\um$, in addition to the ALMA 
photometry outlined above.

The MIR and FIR counterparts of the X-ray AGN in our sample
have already been defined in \cite{Stanley15} using the optical positions of the X-ray AGN 
to match to the following catalogues: 
\textit{Spitzer}--IRAC sources 
as described in \cite{Damen11}, and \cite{Sanders07}, for
GOODS-S and COSMOS, respectively; \textit{Spitzer}--IRS 16$\um$
photometry from \cite{Teplitz11} for GOODS-S; deblended catalogues of MIPS 24$\um$, 
PACS 70$\um$, 100$\um$ and 160$\um$ from 
\cite{Magnelli13}\footnote{The PACS catalogues for and GOODS-S are 
published in \cite{Magnelli13}. The catalogue for COSMOS was created in the 
same way and is available online (http://www.mpe.mpg.de/ir/Research/PEP/DR1).}; 
deblended catalogues of SPIRE 250$\um$, 350$\um$, and 500$\um$ 
 from \cite{Swinbank14}.

All the IRAC catalogues have their detections determined by the 3.6$\um$
maps, the 16$\um$ catalogues and the 24$\um$ deblended catalogues have been produced with the use  
of 3.6$\um$ priors. The PACS and SPIRE deblended catalogues have been produced using the deblended 
24$\um$ catalogues as priors. 
Although in principle we have defined photometry for the full range of 3.6 -- 870$\um$, due to the redshifts
covered by our sample the SED fitting analysis used in our work only makes use of photometry for 24--870$\um$, 
for the majority of the sources.

\subsection{Comparison samples of AGN dominated and star forming galaxies.} \label{comp_sample}
In section~5 we make use of three $z>1$ galaxy samples with published 870$\um$ ALMA photometry, 
in order to explore the F$_{870\um}$/F$_{24\um}$--redshift plane. 
In addition to the X-ray AGN sample of this paper, we use two extreme samples representative of 
AGN dominated 
sources (radio powerful MIR AGN), and star forming galaxies (sub-mm galaxies; SMGs).
Here we provide some more information on these two samples. 

The first comparison sample is that of AGN dominated sources.
The sample consists of AGN dominated galaxies taken from \cite{Lonsdale15}, 
covering the redshifts 0.47 $< z <$ 2.85, and selected to have ultra-red 
{\it WISE} colours and to be radio-loud. 
These are sources lying significantly redward to the main {\it WISE} population in the 
(W1-W2) vs (W2-W3) colour space, where W1 corresponds to 3.4$\um$, W2 to 4.6$\um$, 
W3 to 12$\um$, and W4 to 22$\um$. Samples of sources selected to be the reddest sources 
in the {\it WISE} colour plane have been revealed to be an IR-luminous population of high 
redshift galaxies with strong AGN (e.g., \citealt{Eisenhardt12}; \citealt{Bridge13}; \citealt{Jones14}; \citealt{Tsai15}),
and IR luminosities likely dominated by the AGN emission (e.g., \citealt{Jones15}). 
\cite{Lonsdale15} present ALMA observations and measurements of 870$\um$ of 49 such sources, 
with a resolution of 0.5--1.2$"$, and noise levels of 0.3–-0.6mJy/beam, 
comparable to the ALMA photometry of our sample.
Based on \cite{Lonsdale15}, this sample has AGN bolometric luminosities 
of the order of 10$^{46}\ergss$, covering the high end of AGN luminosities, and has 
been selected to be radio-loud.
Furthermore, \cite{Lonsdale15} estimate the possible contribution from optically thin synchrotron emission to the ALMA 
flux density using multi-frequency VLA data, and argue that none of the sources have strong contamination in their ALMA flux densities.
We use 41 (out of the 49 sources) constrained to redshifts of $z >$1, 
with complementary {\it WISE} photometry. 
The redshifts of the sample are primarily spectroscopic, except for 4 sources with no redshift for which the 
authors assume that $z=2$.

The second comparison sample is that of star forming galaxies, 
and consists of SMGs.
SMGs represent the highly star-forming population at high redshifts, $z\sim$2--3 
(e.g., \citealt{Blain02}; \citealt{Wardlow11}; \citealt{Casey13}), with typical IR luminosities 
of $\rm L_{IR} \sim 10^{46} \ergss$ (e.g., \citealt{Swinbank14}) dominated by emission due to star formation.
The chosen sample of SMGs is taken from the ALMA-LESS survey (A-LESS; \citealt{Hodge13}; \citealt{Karim13}), 
including 122 sources over the redshift range of 0.4 $< z <$ 7 observed with ALMA 870$\um$ during 
Cycle~0. Spectroscopic redshifts where taken from \cite{Danielson17}, photometric redshifts and NIR photometry 
from \cite{Simpson14}, and MIR and FIR photometry from {\it Spitzer}-MIPS 
and {\it Herschel} from \cite{Swinbank14}. In total we use 113 sources of the sample constrained 
to redshifts of 1$<z<$5 (covering a similar redshift range as our sample of X-ray AGN), with spectroscopic redshifts 
for 51 of the sources, the rest being photometric. 
For the ALMA observations of this sample the median resolution was $\sim$1.4$"$, 
and reach typical noise levels of 0.4--0.5 mJy/beam, comparable to the ALMA photometry of our sample.
Although the majority of SMGs is known to be dominated by emission due to star formation, they can still 
be hosts to AGN. \cite{Wang13} presented the X-ray counterparts for part of the A-LESS sample, finding that 8 out 
of the 91 SMGs included, are hosts to X-ray AGN. There have been a number of previous studies identifying AGN in SMG 
samples in both the MIR \citep[e.g.,][]{Valiante07,Pope08, Coppin10} and X-ray 
\citep[e.g.,][]{Alexander05,Laird10}. The X-ray AGN identified in the A-LESS sample 
have X-ray Full Band, 0.5--8keV, luminosities of 10$^{42}$--10$^{44.5}\ergss$ \citep{Wang13}.
We discuss the AGN in this sample further in section 5.2.

\section{IR SED fitting \& decomposition} \label{method}

We performed fitting and decomposition of the IR SED by following and extending 
the methods of \cite{Stanley15}. The SED fitting procedure makes use of a set of
empirical templates describing the IR star formation and AGN emission, 
in order to decompose the SED into the star formation and AGN components.
The set of templates consists of six star-forming galaxy templates and an AGN template (we explore other AGN templates below).
This includes the five star-forming galaxy templates originally defined in \citealt{Mullaney11}, 
(later extended in wavelength by \citealt{DelMoro13}), with the addition of Arp220 from \cite{Silva98}, and 
the mean AGN template defined in \cite{Mullaney11} from a sample of X-ray AGN. 
We asses the impact of our AGN template choice on the SED fitting later in this section, 
and how it compares to other templates in colour-redshift space in Section 5.

Following \cite{Stanley15} we performed two sets of SED fitting to photometry at 8--870$\um$.
The first set includes only the star-forming galaxy templates in the fit, 
while the second set includes both the AGN and star-forming components. 
We fit to the photometric flux density detections, but also force the fits to not 
exceed any of the photometric flux density upper limits.
This procedure results in twelve fitted SEDs to chose from, 
six with and six without the AGN component.
We calculate the integrated 8--1000$\um$ IR luminosity due to star-formation 
from the host galaxy ($\lir$) and due to the AGN ($\liragn$), for each of the twelve fitted SEDs.
To determine the best fitting solution of the twelve fitted SEDs, we
use the Bayesian Information Criteria (BIC; \citealt{Schwarz78}) which allows the 
objective comparison of different non-nested models with a fixed data set. 
The SED fit with the minimum BIC value is defined as the best fit. However, to establish if the 
SED of the source requires an AGN component the SED with the AGN component has 
to have a smaller BIC to that of the SED with no AGN component with a difference of $\Delta$BIC$>$2.
This difference establishes a significant improvement on the fit by the inclusion of the AGN component. 
The uncertainties on the chosen $\lir$, and $\liragn$ values are the combination of the 
formal error on the fit and the range of $\lir$ and $\liragn$ values covered by all template combination
fits with $\Delta$BIC$<$2 to the best fit (see \citealt{Stanley15}).

\begin{figure*}
	\begin{center}
		\includegraphics[scale=0.6]{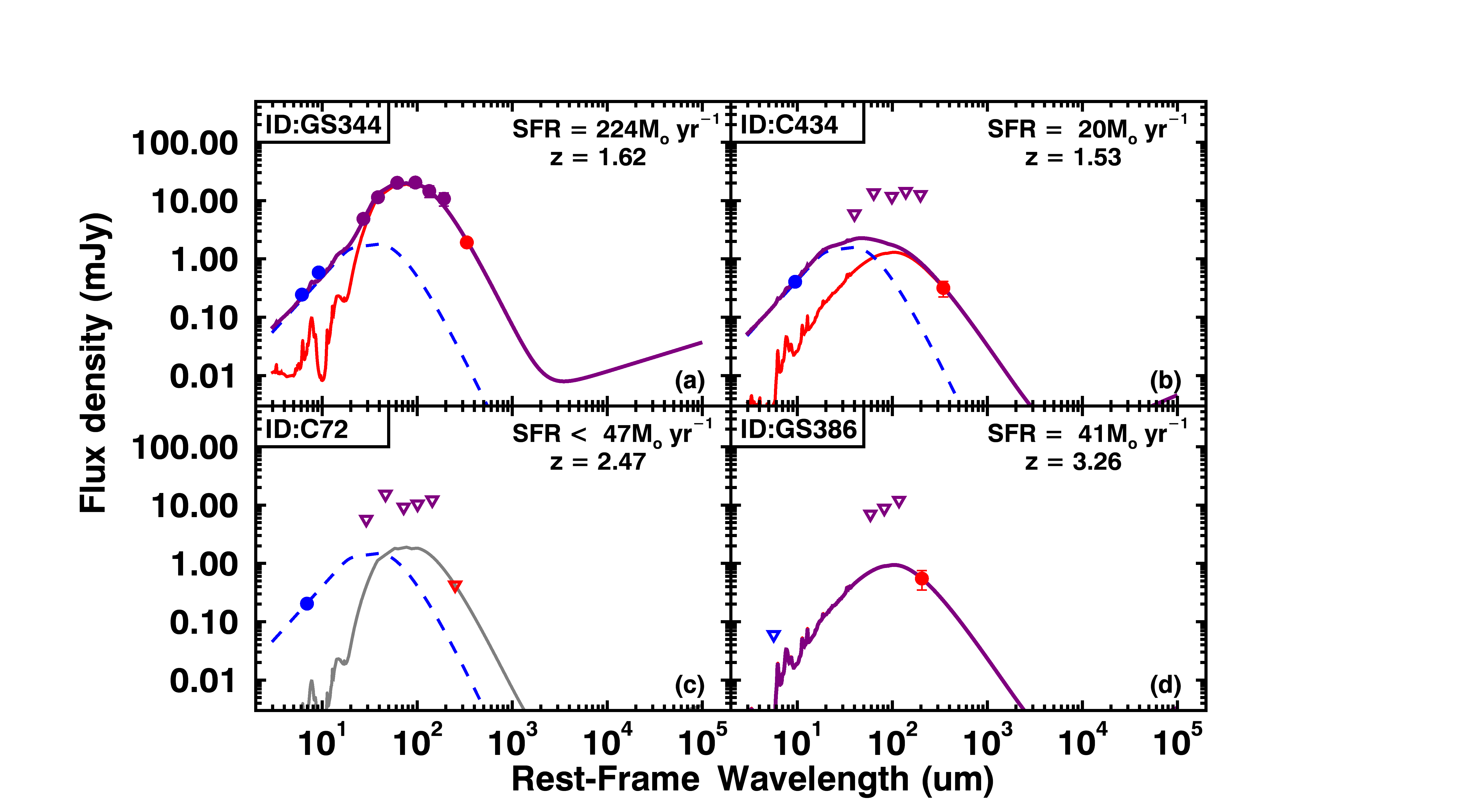}
		\caption[Example SED fits including ALMA photometry.]{Examples of four different cases of SED fitting
			results. In all cases, the blue dashed curve is the AGN component, 
			while the red solid curve is the star-forming component. The total 
			combined SED is shown as a purple solid curve. The grey curves correspond 
			to an upper limit constraint on the SF component.
			The photometry is colour-coded, with blue corresponding to {\it Spitzer} bands, 
			purple to {\it Herschel} bands, and red to the ALMA-870$\um$ photometry. 
			The inverse triangles are upper limits on the flux density. 
			(a) An example where we have full photometric coverage of the SED  (see case (i) in section~\ref{method}). 
			In this case the ALMA photometry on the SED provides confidence in 
			the SED templates used for our analysis. 
			(b) An example where the source is only detected in MIPS-24$\um$ and ALMA-870$\um$ 
			(see case (ii) in section~\ref{method}), and 
			(c) an example of an ALMA undetected source that is only detected in the MIR (see case (v) in section~\ref{method}). 
			In both cases of (b) and (c) the deep ALMA photometry, allow us to constrain 
			the star-forming component to a level that reveals the presence of an AGN 
			component in the MIR. 
			(d) An example were the source is undetected in all bands except 
			for ALMA-870$\um$ (see case (iii) in section~\ref{method}). In the 
			last case we are confident that the emission is dominated by star-formation, 
			as a significant contribution from the AGN the source would result in
			a MIR detection, which is not the case.} \label{fig-seds}
	\end{center}
\end{figure*}

Our fitting results in one of five different situations depending on the number of photometric bands 
a source is detected in.
We detail how we chose the best fit for each below:
\begin{enumerate}
	
	\item If we have more than two photometric detections and at least one is within the 
	FIR range of the rest-frame SED (i.e. at rest-wavelength greater than $\sim$80$\um$ 
	where the peak of star formation emission starts), we are able to decompose the AGN 
	and star formation emission effectively. Therefore, we 
	chose the fit with the minimum BIC value as our best fit. If multiple 
	fits have the same value as the minimum BIC then we take the mean $\lir$, and $\liragn$ of those fits 
	(e.g., Fig.~\ref{fig-seds}(a)). 
	
	\item If a source is only detected in the MIPS--24$\um$ and ALMA--870$\um$ band, we use the 
	comparative BIC values to decide if the IR SED requires the AGN component or not. 
	However, we are unable to discriminate between the different star formation templates. 
	Therefore, we take the mean  $\lir$, and $\liragn$ for the set of fits that best describe the SED 
	(e.g. Fig.~\ref{fig-seds}(b)).
	
	\item If a source is only detected in the ALMA 870$\um$ band we are unable to discriminate 
    between the star formation templates. Therefore, we normalise the star-forming 
	galaxy templates to the ALMA photometry and take the mean of the resulting $\lir$ 
	for the full template range.
	We are confident that if the AGN was significantly contributing to the ALMA photometry, 
	it would have been detected in the MIR at the depth of the MIPS-24$\um$ photometry. 
	Based on the shape of the AGN IR SED, if the AGN was detected at the detection limit of the 24$\um$
	flux density (0.06mJy) it would emit $\sim$6$\times10^{-4}$--0.1mJy at 870$\um$ from redshift 1 to 
	4.7 respectively. The highest contribution possible by the AGN to the 870$\um$ flux density, for the sources
	in our sample, would be at a redshift 4.7, and would 
	only account for $\sim$6\% 
	of the measured flux density of the source at that redshift. An example of this is given in
	Fig.~\ref{fig-seds}(d) where we show the case of a $z=3.26$ galaxy detected only at 870$\um$. If the 
	AGN was to emit the observed 870$\um$ flux density of 0.4mJy then the 24$\um$ flux density should be 
	$\sim$4mJy, a value significantly larger to that of the flux density limit.
	
	\item If a source has only MIR detections, or no detections at all, then we cannot confidently decompose
	the SED and so we constrain an upper limit on the star-forming component using the limits and/or 
	the 3$\sigma$ error on the detections.
	We normalise all star-forming templates to the lowest value of the upper limits, 
	including as a limit the 3$\sigma$ above the photometry if the source is detected in a given MIR band. 
	We then take the maximum $\lir$ value of the range of normalised templates, as the upper limit. 
	The same is done for the estimation of the $\liragn$ upper limit.
	
	\item If a source is detected in the MIR and the limit on the star-forming component (constrained by the limits at $>$80$\um$) is 
	$>$5$\sigma$ below the observed frame 8--24$\um$ photometry, then we can identify the
	presence of an AGN component. We find that in these cases we can measure the $\liragn$, 
	even if we can only constrain an upper limit on the $\lir$ (e.g., Fig.~\ref{fig-seds}(c)). 
	
\end{enumerate}

Following this method, we have performed individual SED fitting for 
the whole sample of X-ray AGN studied here. The results from the SED fitting 
procedure are given in Tables \ref{tabresults_gs} \& \ref{tabresults_c} along 
with the X-ray hard band luminosity ($\lx$) and redshift ($z$) of the sources.
The best-fit SEDs for all sources are given in Fig.~\ref{bestfits}. 
Interestingly, where we could only identify a MIR AGN 
component in 1 of our sources prior to ALMA observations, 
we can now confidently identify a MIR AGN component in 54/109 ($\sim$50\%) of 
the ALMA observed sample, with AGN fractions down to 20\% of the total IR (8--1000$\um$) luminosity. 
Throughout this paper we only consider that a source has a MIR AGN component in their SED when the 
fit requires an AGN component with a significant contribution (at least 20\%), while SED fits that require an AGN
component with a very weak contribution (less than 20\%) are considered uncertain. These 
sources are flagged in Tables \ref{tabresults_gs} \& \ref{tabresults_c}, with a flag of 1 for 
weak/uncertain AGN components 
in the fit, and a flag of 2 for fits with a significant AGN component.

We note that a comparison between the observed $\lx$ values and the measured 
6$\um$ luminosities from the AGN component of our SED fits (when present), shows a good agreement between the 
two. Specifically, the majority of the sources with an AGN component in their SED fits lie within the scatter of the local AGN relation (e.g., \citealt{Lutz04}). There is one source lying significantly offset from the local relation. This source has an observed $\lx$ value lower than the 6$\um$ luminosity by 1.6dex (factor of $\sim$40), which is consistent with the measured column density of N$_{\rm H}=9\times10^{23} \rm cm^{-2}$ (from \citealt{Luo17}).

We have followed the same SED fitting method for the two comparison samples described in section~2.3, 
using the available published photometry.
Overall, with our SED fitting procedure we have an $\lir$ measurement for 21/41 (51\%) of 
the AGN dominated sources with the rest having a well constrained upper limit. As expected, 
we identify an AGN component
in all 41 of the AGN dominated sources with a minimum AGN contribution to the total IR luminosity 
of 50\%, and with 22/41 (54\%) of the sample having an AGN component that contributes $\gtrsim$90\% 
of the IR luminosity. When looking at the star forming galaxy sample, our SED fitting process can constrain an $\lir$ 
measurement for the whole sample, and finds that all of the sources have IR emission dominated by 
star-formation, with only 12/113 (11\%) of the sources having an identified IR AGN component. 
The $\lir$ values of these comparison 
samples cover the range of $\sim$0.2--3$\times10^{47}\ergss$ and 
$\sim$0.2--4$\times10^{46}\ergss$ for the AGN dominated and star forming galaxies, 
respectively (see also tables A3 and A4).

In our analysis we have only used one AGN template, that of \cite{Mullaney11} defined for 
a sample of nearby X-ray AGN. However, there is a number of other AGN templates defined for 
different samples (e.g., \citealt{Mor12}; \citealt{Symeonidis16}; \citealt{Lani17}; \citealt{Lyu17}). 
Since many of our sources are found to have a strong AGN 
component in their IR SED, we need to test if the results are affected by the choice of AGN template. 
The most deviant AGN template from our primary choice is that of \cite{Symeonidis16}, defined for a sample of
optical PG quasars. This template can have a stronger IR contribution than that of \cite{Mullaney11},
due to the fact that it is characterised by a more gradual drop-off at long wavelengths. However, 
recent work by \cite{Lani17} and \cite{Lyu17}, have demonstrated that for the same or similar samples of 
PG quasars the AGN template derived is actually much more similar to that of \cite{Mullaney11}, 
than \cite{Symeonidis16}, shedding some uncertainty on the later template. 
Furthermore, when we examine the F$_{870\um}$/F$_{24\um}$ -- $z$ plane in section~5, we 
demonstrate that the \cite{Symeonidis16} template is inconsistent with the colours of most AGN dominated sources.
Finally, using the AGN templates with a steeper drop-off at the longer wavelengths, has a minimal effect on our 
derived SFRs, typically at only a few \% level (see \citealt{Stanley15,Stanley17}).

\section{Improvements on $\lir$ constraints}\label{improvements}
To demonstrate how much better we can constrain $\lir$ for our 
sample once we have ALMA photometry in addition to {\it Spitzer} and 
{\it Herschel}, we have performed the same SED 
fitting analysis on the sample with and without the ALMA photometry.
Here we quantify the improvements achieved on the $\lir$ values.

\begin{figure*}
	\begin{center}
		\subfloat{\includegraphics[scale=0.365]{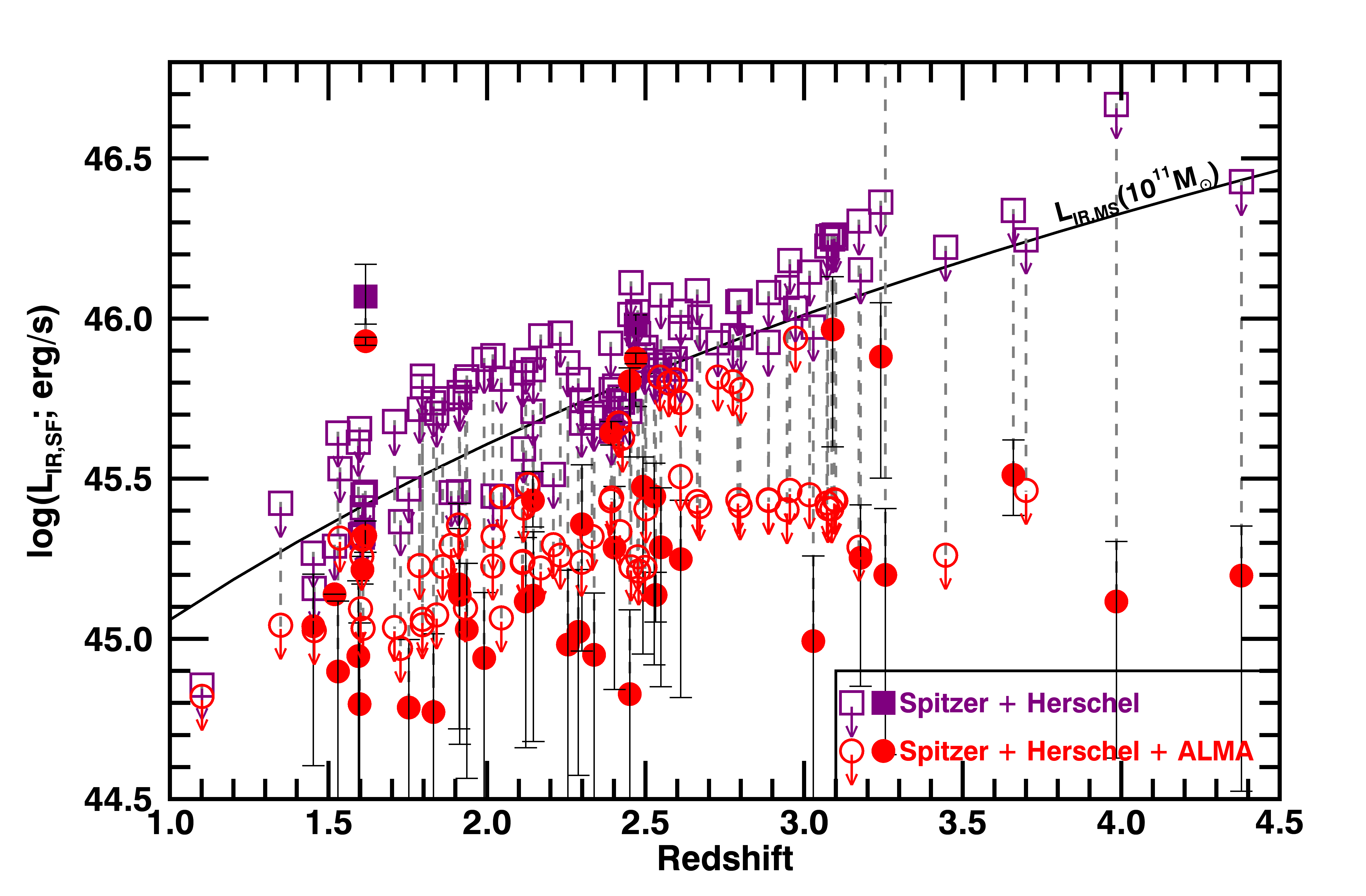}}
		\subfloat{\includegraphics[scale=0.365]{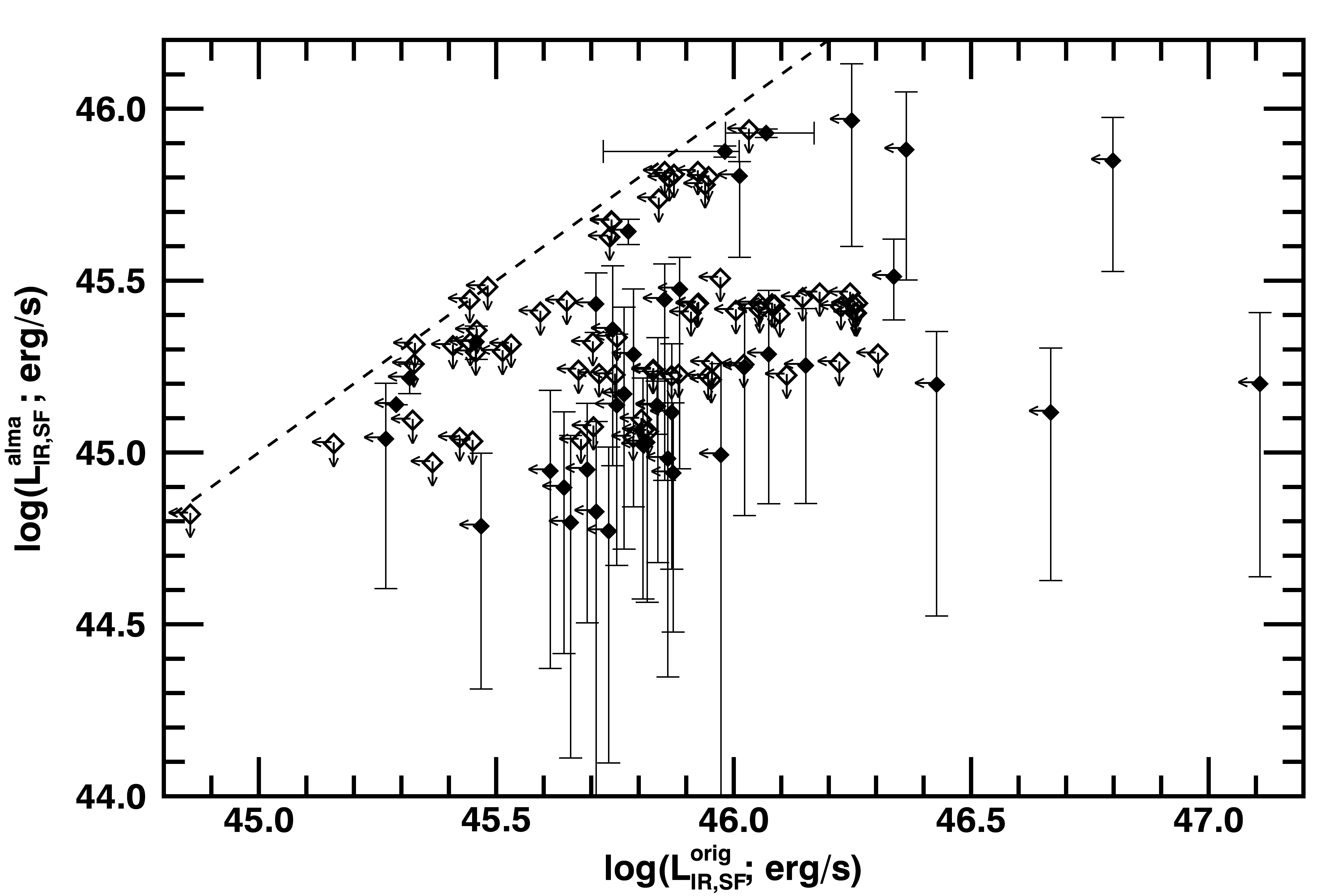}}
		\caption{{\em (left)} IR luminosity due to star formation ($\lir$) as a function of redshift for our sample
			before (purple) and after (red) the inclusion of deep ALMA photometry in our SED fitting. 
			 {\em (right)} IR luminosity due to star formation after the inclusion of the ALMA photometry ($\lir^{alma}$) 
			as a function of the IR luminosity due to star formation before the inclusion of the ALMA photometry ($\lir^{orig}$), 
			with the dashed line corresponding to the 1 to 1 ratio.
			We now have 20 times more measurements than previously possible, with 
			 40/109 sources having an $\lir$ measurement. For 73/109 (67\%) of the sources the 
			measurements and upper limit constraints on $\lir$ have typically decreased by factors of 2--10 compared to the original upper limit constraints (see Fig~4).} \label{fig-baa}
	\end{center}
\end{figure*}
\begin{figure}
	\begin{center}
		\includegraphics[scale=0.5]{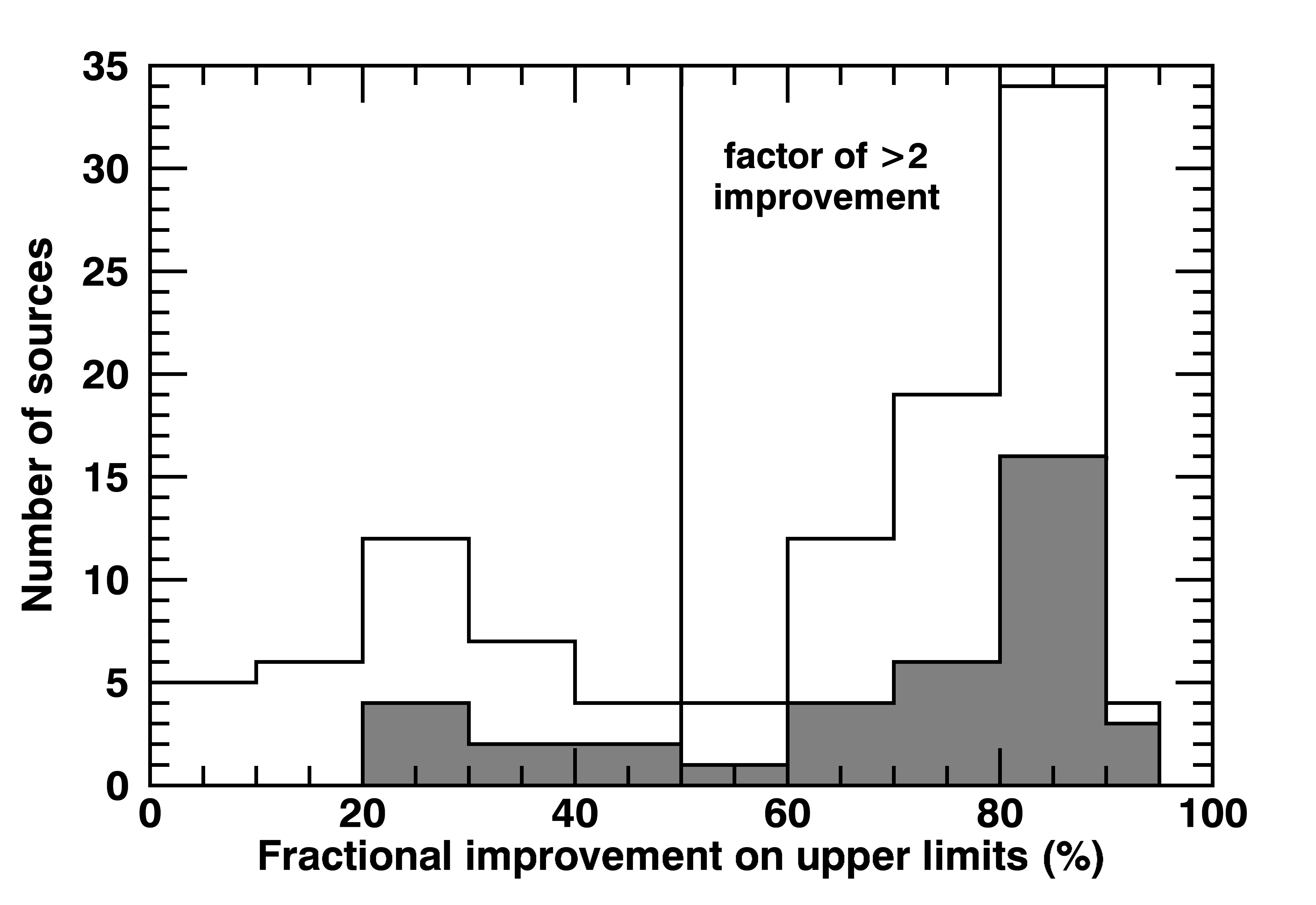}
		\caption{Histogram of the improvement in the $\lir$ values and upper limits when ALMA photometry was included ({\em alma}), compared to the original upper limit constraints ({\em orig}), 
		i.e. $(\lir^{orig}-\lir^{alma})/\lir^{orig}$. Filled in black are the values corresponding to sources that turned from upper limits in the original fit, to measurements when including the ALMA photometry.
			The majority of the upper limits in our sample (73/109) have new $\lir$ measurements or upper limits that have decreased by more than a factor of 2. We note that from the remaining 36/109 sources that have less than a factor of 2 improvements, 14 where observed in the wrong configuration. As a consequence these 14 sources have larger RMS values than the rest of the sample (see section~2.1; Scholtz et al. 2018).
		} \label{fig-imphist}
	\end{center}
\end{figure}

In Fig.~\ref{fig-baa}{\it (left)} we show $\lir$ when constrained using 
8 -- 500$\um$ photometry (purple) and 8 -- 870$\um$ (red) 
photometry (i.e., without and with the ALMA photometry), 
as a function of redshift. For comparison we also plot the track for the mean 
$\lir$ of star-forming main sequence galaxies (e.g, \citealt{Schreiber14}) with stellar 
masses of M$_* = 10^{11} \Msol$ (the rounded median stellar mass for our sample; 
Scholtz et al. 2018).
In Fig.~\ref{fig-baa}{\it (right)} we show a different representation of the comparison, 
by plotting the new $\lir$ values constrained with the additional 
ALMA photometry ($\lir^{alma}$) against original $\lir$ values constrained 
without the ALMA photometry ($\lir^{orig}$). With a dashed line we give the 
1 to 1 ratio.
As the sample was selected to be {\it Herschel} undetected/FIR faint, the majority 
of the sources (107/109; 98\%) only had upper limit constraints on their $\lir$ 
values in the absence of the ALMA 870$\um$ photometry. 
The ALMA photometry allows us to both measure the $\lir$ of sources not possible previously 
(from 2\% to 37\% of the sample), and to also push the limits on $\lir$ values to 
significantly lower 
levels, up to a factor of 10 (see Fig. \ref{fig-baa}, and \ref{fig-imphist}). 
For sources with 
sufficient {\it Herschel} constraints to measure $\lir$ (2/109) we find a change 
in $\lir$ when including the ALMA photometry of only a factor of 1.3 and 1.4. 
The agreement of the ALMA photometry to the {\it Herschel} constraints provides extra 
confidence in our SED fitting approach and choice of templates, even in the absence of 
ALMA photometry. 

In Fig.~\ref{fig-imphist} we show a histogram of the improvement in constraining the 
$\lir$ values of the 98\% of our sample that originally only had upper limit constraints 
based on {\it Herschel} photometry. The value plotted is given by the equation: 
$\frac{\lir^{orig}-\lir^{alma}}{\lir^{orig}}$. With the filled regions of the histogram we 
highlight the sources that turned from upper limits to measurements. 
It is immediately clear that more than half of our sample (67\%) have $\lir$ constraints that have changed 
by more than a factor of 2. The apparent bi-modality in the improvements of the upper limit constraints is 
driven by the range of RMS values for our observations. For the subsample of 14 sources incorrectly observed 
with high resolution, the resulting RMS of the heavily tapered ALMA maps is as high as 0.8mJy/beam, which 
results in only a small improvement on the constraints of the $\lir$ upper limits (see section~2.1; \citealt{Scholtz18}).

Overall, we now have $\lir$ measurements for 40/109 ($\simeq$37\%) of the sources, 
that is 20 times more sources than what was possible without the ALMA photometry. 
For the sources that still have an upper limit constraint (69/109; $\simeq$63\%), 
the values have lowered by up to a factor of 10 with the addition of ALMA data. 
Furthermore, the majority of our sample (67\%) have improved by more than a factor of 2, 
and we can now identify an AGN component in the IR SEDs of 50\% of our sample
compared to the original 0.9\%. 
In summary, we have demonstrated that deep ($\sim$0.1--0.3mJy) 870$\um$ ALMA observations, 
in combination with {\em Spitzer} and {\em Herschel} data, significantly improve the AGN-star formation 
SED decomposition and SFR measurements for distant X-ray AGN. Such improvements make it possible 
to constrain SFR distributions of this population rather than just investigate mean properties (Mullaney et al. 2015; Scholtz et al. 2018).

\section{The AGN IR emission: Identifying AGN through their F$_{870\um}$/F$_{24\um}$ ratio} \label{agn_comp}
With the excellent constraints on the star formation component of the IR SED that the ALMA observations can provide, 
we are now able to better constrain the MIR emission of the AGN itself. The shape of the star-forming IR SED, 
in combination with the constraints placed on it by the ALMA 870$\um$ photometry, allows for the detection 
of a MIR excess, even when a source is undetected at 870$\um$. Indeed, as mentioned in section~3 we 
can now confidently identify a MIR AGN component in $\sim$50\% of 
the ALMA observed sample, with AGN fractions down to 20\% of the total IR luminosity. 
 
The deepest data in the extragalactic deep fields, such as CDF-S and COSMOS, 
within the wavelength range of the IR SED are from 24$\um$ 
({\it Spitzer}-MIPS) and 870$\um$ (ALMA Band-7) observations. For a composite source, that 
has both AGN and star formation emission in the IR, having detections and/or deep upper limits
of the flux density at those wavelengths may allow for a successful decomposition of the AGN and SF components.
For this reason we explore the parameter space of the ratio of the flux densities at 870$\um$ over 
24$\um$ as a function of redshift, for the potential of identifying 
AGN dominated and composite sources. Throughout the rest of this paper we call this the F$_{870\um}$/F$_{24\um}$-redshift plane, 
where F$_{870\um}$ is the flux density of the ALMA Band-7 at 870$\um$ and F$_{24\um}$ is the flux density of the {\it Spitzer}-MIPS 24$\um$ band.
In order to do this we use three different samples: (1) the X-ray AGN sample of this study 
that mostly contains composite sources; (2) an AGN dominated galaxy sample; and (3) a star forming galaxy sample (see section~2.4). 
We have chosen the two additional samples in order to cover the two extremes of AGN dominated IR SEDs, 
and star formation dominated IR SEDs, as well as the range of composites between them. Samples (2) \& (3) 
are described in section 2.4.

In section~5.1 we use the SED templates for the AGN and SF components in order to define 
the F$_{870\um}$/F$_{24\um}$-redshift plane, and use the three galaxy samples to verify the 
AGN, star formation, and composite regions. In section~5.2 we compare the selection of AGN candidates
based on the F$_{870\um}$/F$_{24\um}$ ratio, to the findings from our SED fitting analysis, and to existing IRAC
colour selection criteria.

\subsection{Defining the F$_{870\um}$/F$_{24\um}$-redshift plane for infrared AGN identification studies.}

We define the regions of the (F$_{870\um}$/F$_{24\um}$)-redshift 
plane dominated by purely AGN emission and by purely star-forming 
emission using the star-forming templates of our SED fitting procedure, 
and the AGN templates of \cite{Mullaney11}, the mean of which is used in our 
SED fitting procedure (see section~3). 
For comparison and to explore the (F$_{870\um}$/F$_{24\um}$)-redshift 
plane, we also include an additional two AGN templates, 
and an additional set of SF templates.
We use the set of star-forming templates from \cite{Dale02} produced by a phenomenological model of 
star-forming galaxies, and the AGN templates of \cite{Mor12} and \cite{Symeonidis16} derived for 
samples of luminous quasars that cover the extremes in FIR/MIR colours for AGN templates from the literature. 

\begin{figure*}
	\begin{center}
		\includegraphics[scale=0.5]{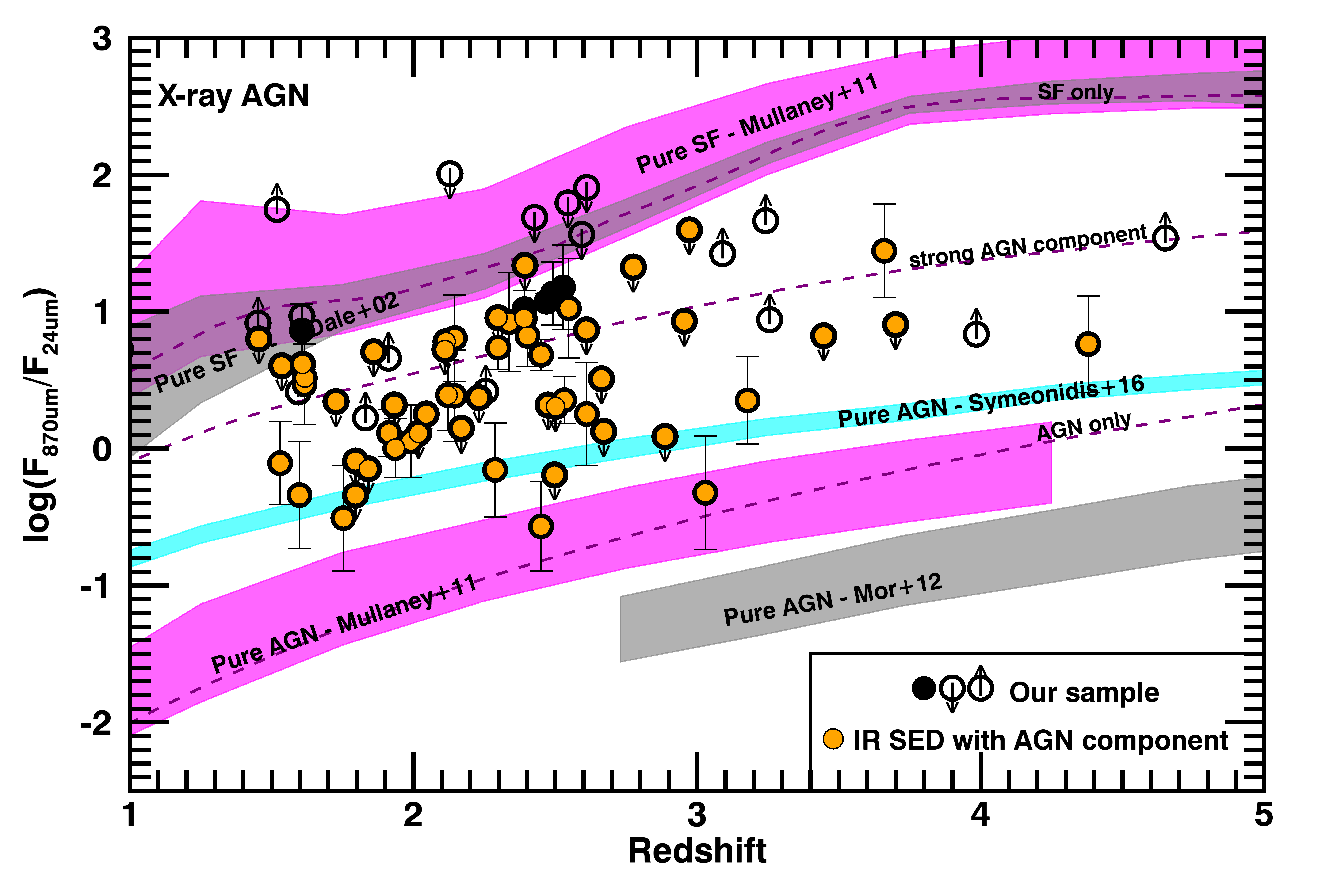}
		\caption{The F$_{870\um}$/F$_{24\um}$ ratio as a function of redshift for the ALMA observed sample of X-ray AGN. 
			Sources for which our SED fitting finds an AGN component with more than 20\% 
			contribution to the IR emission are highlighted with yellow centres. 
			The purple dashed curves correspond to the median F$_{870\um}$/F$_{24\um}$ ratio as a function of redshift for SEDs 
			with 0\% AGN contribution (SF only), 50\% AGN contribution (strong AGN component), and 100\% AGN contribution (AGN only) 
			to the IR luminosity.  
			} \label{fig-f24f870}
	\end{center}
\end{figure*}

\begin{figure}
	\begin{center}
		\includegraphics[scale=0.37]{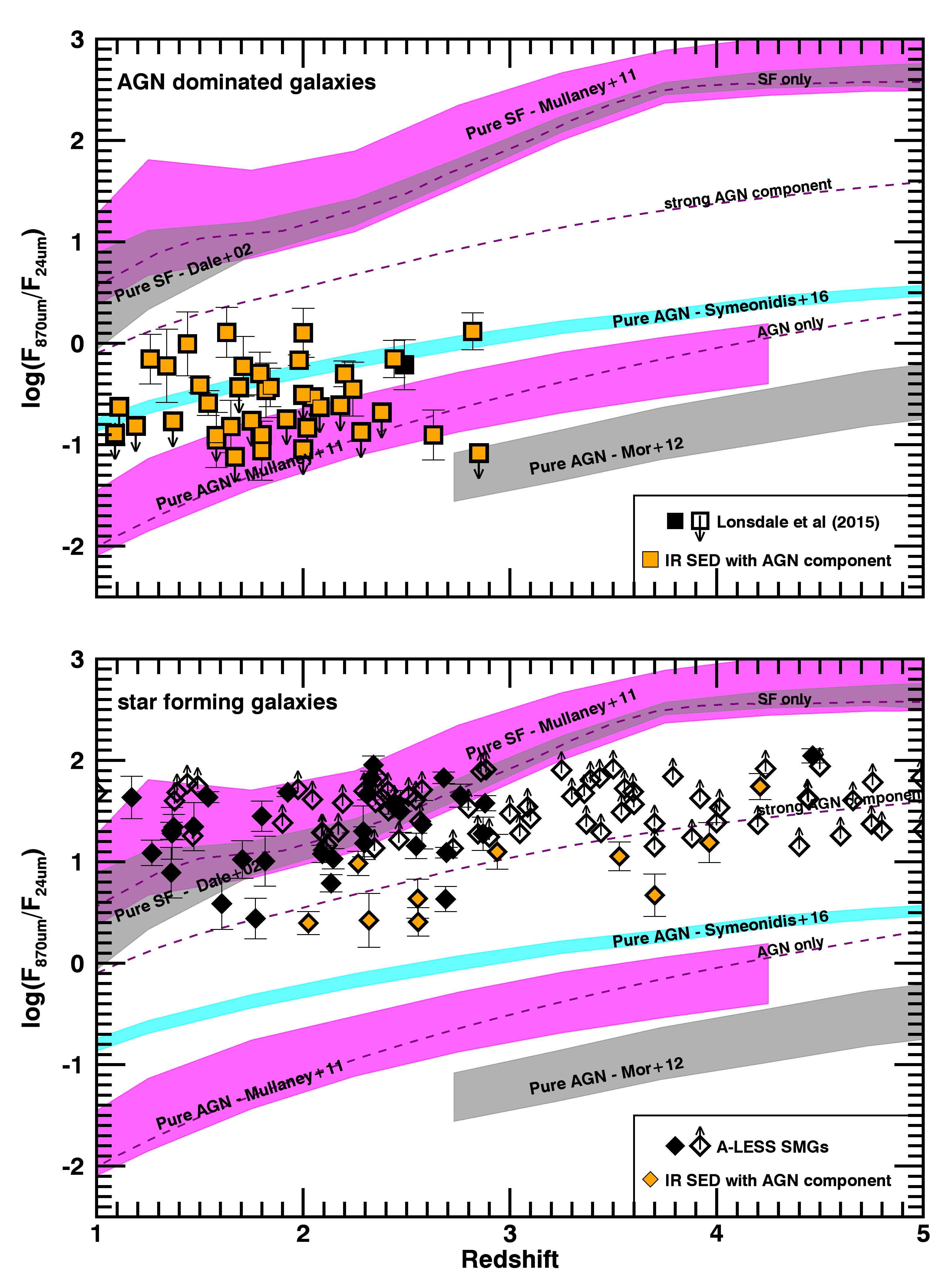}
		\caption{The F$_{870\um}$/F$_{24\um}$ ratio as a function of
			redshift, for two comparison samples also observed with ALMA. 
			({\it top}) The sample of AGN dominated sources from \citealt{Lonsdale15}.				
			({\it bottom}) A sample of SMGs from the A-LESS survey (e.g., \citealt{Hodge13}). Sources for which 
			our SED fitting finds an AGN component with at least 20\% 
			contribution to the IR emission (flagged with 2 in Tables~\ref{tabresults_gs}\&\ref{tabresults_c}), are highlighted with yellow centres.
			The purple dashed curves correspond to the median F$_{870\um}$/F$_{24\um}$ ratio as a function of redshift for SEDs 
			with 0\% AGN contribution (SF only), 50\% AGN contribution (strong AGN component), and 100\% AGN contribution (AGN only) 
			to the IR luminosity.  
			The strong AGN component curve seems to separate well the F$_{870\um}$/F$_{24\um}$-redshift plane 
			in the two regions covered by star forming galaxies and AGN dominated sources. For this reason we test it 
			as an AGN candidate selection limit in section~5.2.} \label{fig-f24f870_comp}
	\end{center}
\end{figure}

We plot the (F$_{870\um}$/F$_{24\um}$)-redshift plane for the three different samples in 
Fig.~\ref{fig-f24f870}~\&~\ref{fig-f24f870_comp}. With coloured regions we show the parameter 
space covered by the star-forming templates (the \citealt{Mullaney11}+Arp220 set of 
templates in pink; the set of templates from \citealt{Dale02} in grey), and the 
region covered by AGN templates (from \citealt{Mullaney11} in pink; \citealt{Mor12} in grey; and 
\citealt{Symeonidis16} in blue). We note that the template of \cite{Mor12} is limited to redshifts of 
$z\geq2.7$ in our plots, due to the restricted wavelength region (0.5--250$\um$) it has been defined for.
There is a clear divide between the regions covered by the star formation and AGN templates. 
This is due to the relative shapes of the AGN and star formation IR SEDs 
(see blue dashed, and red solid curves in Fig.~2), which results in sources with a significant contribution 
from the AGN component having a 24$\um$ flux density dominated by the AGN emission, while the 870$\um$
flux density will be dominated by the star-formation (except for cases of pure AGN emission).

When plotting the X-ray AGN sample that consists of a wide range of AGN -- SF composite sources, 
it covers the full range between the star formation and AGN region of the plane 
(see Fig.~\ref{fig-f24f870}). This is not surprising as the X-ray sample covers a broad range of 
X-ray luminosities, and there can be a wide range of SFR values for a fixed AGN 
luminosity in samples of X-ray AGN (e.g, \citealt{Mullaney15}; see section 4.3 of \citealt{Stanley15}). 
To test if the star formation and AGN regions of the plane are indeed 
representative of star forming galaxies and AGN dominated sources, we use 
the two samples described in section 2.4, one representative of AGN dominated sources, 
and one representative of star forming galaxies. In Fig.~\ref{fig-f24f870_comp} we plot the 
F$_{870\um}$/F$_{24\um}$-redshift plane for these two samples. 
The AGN dominated sample lies at F$_{870\um}$/F$_{24\um} <$1.6 and towards the AGN region of the plane.
The star forming galaxy sample lies at F$_{870\um}$/F$_{24\um} >$1.6 and towards 
the star formation region of the plane. The agreement between the colours of the AGN dominated, 
and star forming galaxies and our templates 
is an additional indication for their suitability for our SED fitting analysis.

We compare the AGN dominated sample to the regions of the plane covered by the different AGN templates, in order 
to asses how compatible or incompatible these AGN templates are with the observed F$_{870\um}$/F$_{24\um}$.
Sources with F$_{870\um}$/F$_{24\um}$ ratios on and above those of an AGN template are 
considered compatible with it, while sources with F$_{870\um}$/F$_{24\um}$ ratios below those of the AGN template
are incompatible. This is due to the fact that a F$_{870\um}$/F$_{24\um}$ ratio below that of a given AGN template simply 
cannot be described by that template, while a F$_{870\um}$/F$_{24\um}$ ratio above can be described as a composite of the AGN template
and star formation emission. 
We find that the AGN template of \cite{Mullaney11} is compatible with 40/41 sources, the \cite{Mor12} template is likely 
compatible with all 41. \footnote{Due to the truncation of the \cite{Mor12} template at 250$\um$ we only calculate the 
F$_{870\um}$/F$_{24\um}$ ratio 
from redshifts $z>2.7$ (plotted with a grey region in Fig.~5--7). However, with simple extrapolation to lower redshifts we can expect 
that all AGN dominated sources are compatible with the template.} In contrast, 28/41 sources lie below the F$_{870\um}$/F$_{24\um}$ ratios of the \cite{Symeonidis16} AGN template, by an average factor of $\sim$2.
Consequently, the \cite{Symeonidis16} AGN template is the most incompatible to 
the F$_{870\um}$/F$_{24\um}$ ratios of the AGN dominated sample.

To further quantify the location of AGN candidates in the F$_{870\um}$/F$_{24\um}$-redshift plane we make use
of the SED templates used in our SED fitting analysis (see section~3), in order to create composite SEDs with a specific AGN contribution.
We use these to define the expected F$_{870\um}$/F$_{24\um}$ ratio as a function of redshift for composites with 
a strong (50\%) AGN contribution to the IR luminosities, in order to distinguish different 
AGN contributions within the composites region of the F$_{870\um}$/F$_{24\um}$-redshift plane. 
We combine each SF template to our AGN template to create composite SEDs with a 50\% AGN 
contribution to the total IR luminosity. We then take the median composite SED. 
This SED is then shifted with redshift steps of 0.2 from redshifts 1 to 5, and at each step we calculate the 
observed frame F$_{870\um}$/F$_{24\um}$ ratio. As a result we have a measure of the median F$_{870\um}$/F$_{24\um}$ ratio as a function of redshift, for SEDs with a strong AGN component. 
We show the expected F$_{870\um}$/F$_{24\um}$ ratio as a function of redshift for 
sources with a 50\% AGN contribution to the IR luminosity in Fig.~5--7, with a dashed purple track labelled ``strong AGN component", which follows the form:
\begin{equation}
 \log_{10}(\frac{F_{870\um}}{F_{24\um}}) \, = \, -1.19 + 3.623\times \log_{10}(1+z) 
\end{equation}
 In Fig.~\ref{fig-f24f870}~--~\ref{fig-irac} we also 
show the median  F$_{870\um}$/F$_{24\um}$ ratio as a function of redshift for only the star formation components, and 
for only the AGN component, with dashed purple tracks labelled ``SF only" and ``AGN only" respectively. 

The track defined for a strong AGN component seems to discriminate well between the two samples of AGN and star forming galaxies, 
except for 8 sources of the star forming galaxy sample (see Fig.~\ref{fig-f24f870_comp}). These 8 sources appear to have 
AGN signatures at various wavelengths (see section 5.2 for further 
discussion). In the next sub-section we discuss the potential of using Eq.~1 as a method for identifying sources with strong 
MIR AGN emission. 

\subsection{AGN identification: Application of a F$_{870\um}$/F$_{24\um}$ selection and comparison to other approaches}
The strong AGN component line defined in the previous section does a good job of discriminating between AGN dominated 
and SF dominated samples (see Fig.~\ref{fig-f24f870_comp}), and can select composite sources with 
a strong ($>$50\%) AGN contribution to the IR emission. Here we use Eq.~1, that describes the strong AGN component
line, as the F$_{870\um}$/F$_{24\um}$ selection limit for AGN candidates, and compare to MIR selection methods 
(e.g., \citealt{Stern05}; \citealt{Donley12}) and the results of our SED fitting analysis. 
We note that the following discussion is limited to X-ray AGN that are {\em Herschel} faint or non-detected 
based on our sample selection (section~2.1). This sample selection may contribute to the low number of sources detected in all four 
IRAC bands (50/109; 46\% of the sample). In order to do the comparison to the MIR colour selection, we restrict our X-ray AGN 
sample to only those 50 sources. We also restrict the star forming galaxy sample to 81/113 sources detected in all four IRAC bands.

\begin{figure*}
	\begin{center}
		\subfloat{\includegraphics[scale=0.55]{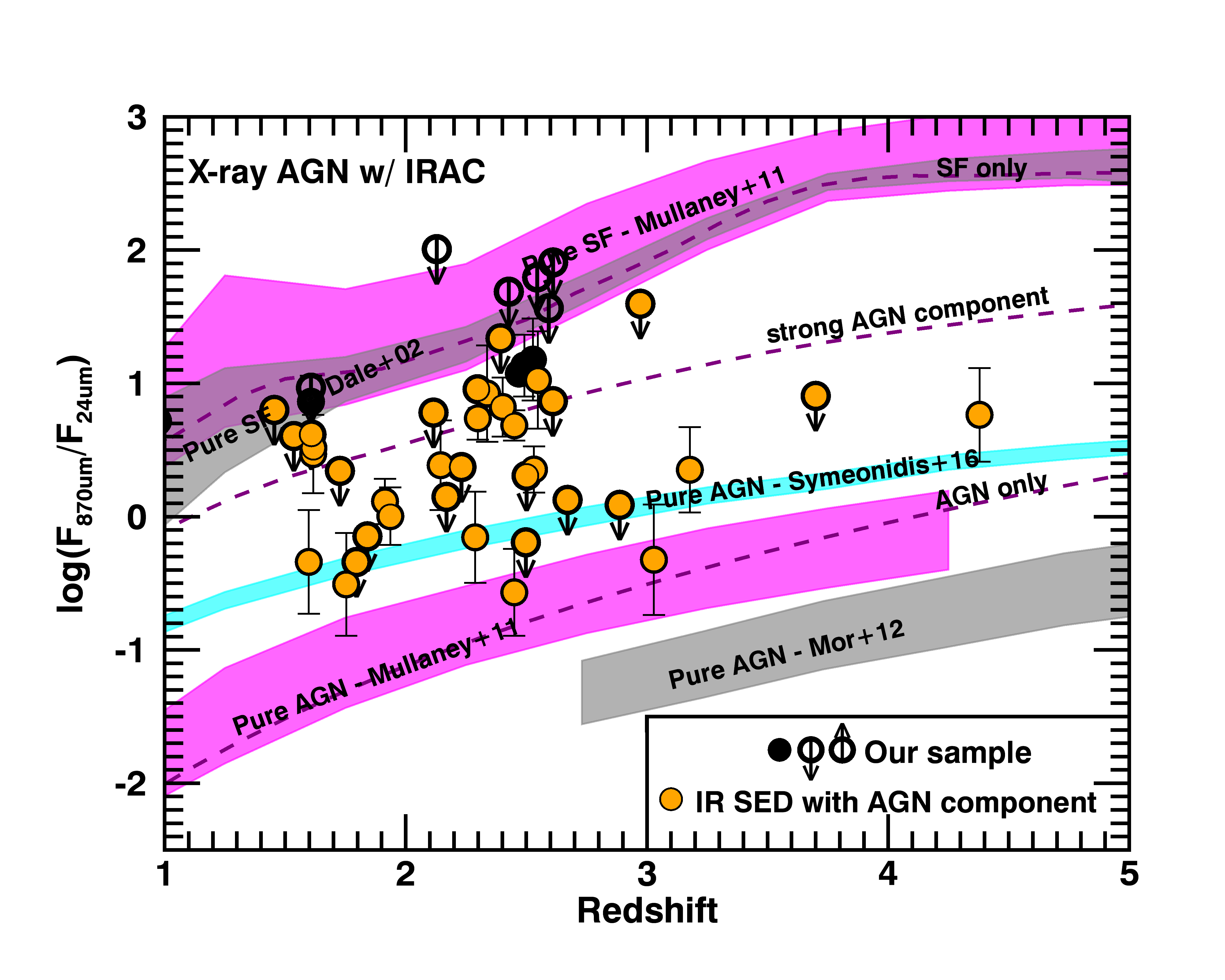}}
		\subfloat{\includegraphics[scale=0.55]{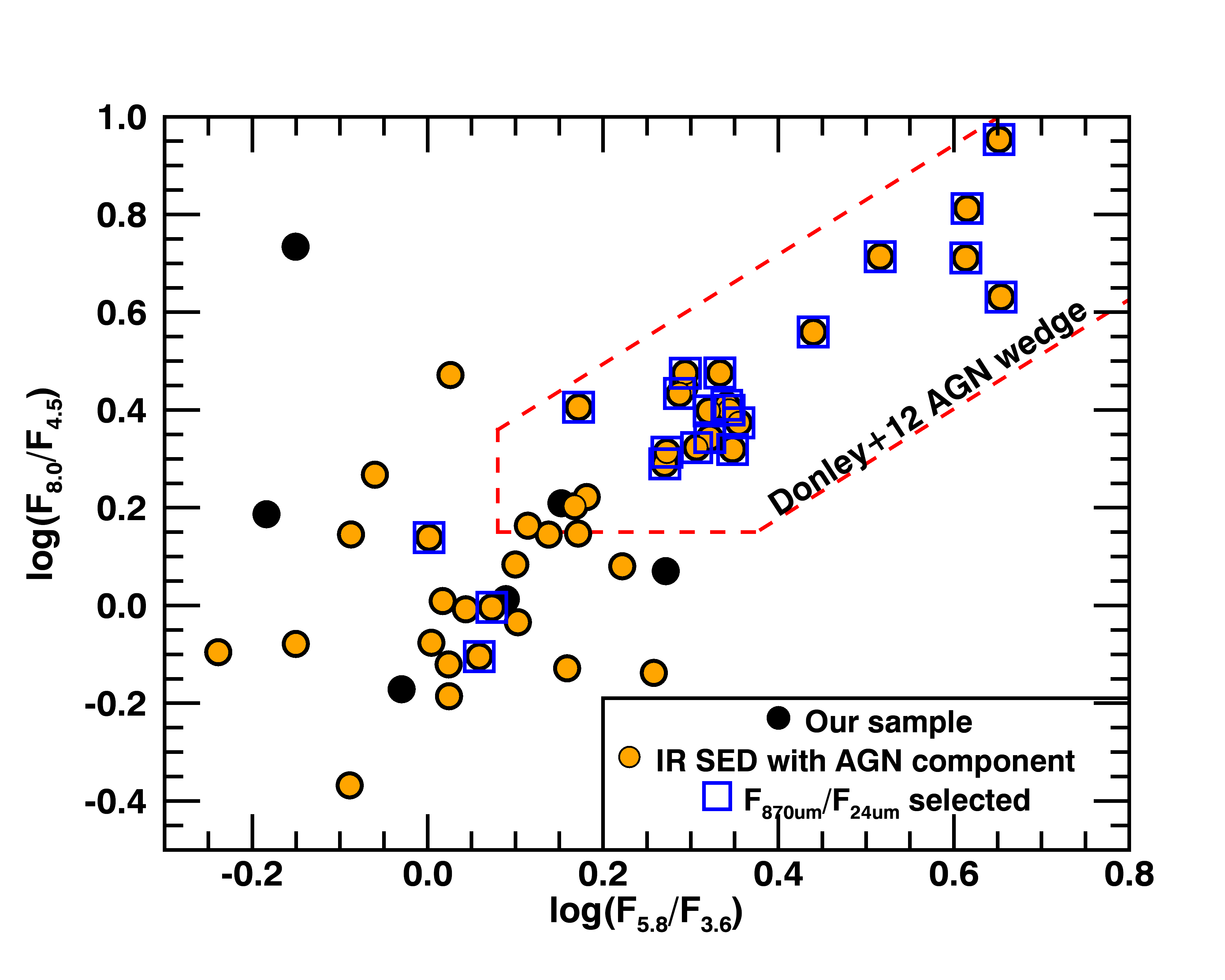}}
		\caption{Comparison of the F$_{870\um}$/F$_{24\um}$ ratio selection to the commonly used IR colour selection
			for the 50 X-ray AGN that are detected in all four IRAC bands. 
			({\em left}) The F$_{870\um}$/F$_{24\um}$ ratio as a function of redshift. 
			Using the ``strong AGN component" line (dashed purple curve) as a selection limit for AGN candidates, 
			we find that 22/50 sources are selected.
			({\em right}) The colour-colour diagram based on IRAC photometry for our sample of X-ray AGN.
			The AGN selection criteria of Donley et al. (2012) are shown with the dashed lines, with sources within 
			the enclosed area being AGN. The sources that are selected as AGN candidates from 
			the ``strong AGN component" F$_{870\um}$/F$_{24\um}$ line are highlighted with a blue square.
			We find that the majority of the F$_{870\um}$/F$_{24\um}$ selected AGN candidates (19/22) 
			are also selected by their IRAC colours.
			In both plots, we indicate the sources where the best fitting solution from the 
			SED-fitting requires an AGN contributing $\gtrsim$20\% with yellow centres. 
			The sources with weaker AGN components (contributing $\approx$20--50\% of the IR luminosity), can be missed by both colour criteria.
			} \label{fig-irac}
	\end{center}
\end{figure*}

In the case of our X-ray AGN sample, the F$_{870\um}$/F$_{24\um}$ limit selects
22/50 sources as AGN candidates. Of these 22 sources, all have a strong AGN component in 
their best-fit SEDs. To see how many would be selected by the more commonly used MIR colour selection, 
we use the \cite{Donley12} IRAC colour criteria for identifying MIR AGN, 
that have the lowest contamination from non-AGN sources compared to previous IRAC selection criteria (e.g., \citealt{Stern05}).
The IRAC colour criteria select 19 out of the 22 sources selected by the F$_{870\um}$/F$_{24\um}$ limit.
In Fig.~\ref{fig-irac} we show the two selection methods, with the F$_{870\um}$/F$_{24\um}$--$z$ plane is shown in Fig.~\ref{fig-irac}{\it (left)},
and the IRAC colour--colour plane in Fig.~\ref{fig-irac}{\it (right)}. We note that there are 5 sources selected by the IRAC colour criteria, that
have not been selected by the F$_{870\um}$/F$_{24\um}$ limit. This is due to the fact that these sources have AGN components with a contribution of 0--47\% to the total IR luminosity, and by definition the F$_{870\um}$/F$_{24\um}$ limit discussed here will select only sources with $>$50\% AGN contribution.
Overall, both methods are comparable in selecting source with a strong AGN component, but both will miss the majority of sources that have
AGN components contributing $<$50\% to the total IR luminosity.

In the case of the AGN dominated sample, the F$_{870\um}$/F$_{24\um}$ selection limit
successfully selects the full sample of  41 sources. These sources have been selected through 
their {\it WISE} colours, and so all of them are already IR colour selected, and all 41 sources have a strong AGN component in 
their best-fit SEDs. 

In the case of the star forming galaxy sample 
the F$_{870\um}$/F$_{24\um}$ selection limit selects 8/81 sources as having a strong AGN component. 
Of these 8 sources, 7 have a confident AGN component in their best-fit SEDs (contributing 30--83\% to the IR luminosities),
and 1 would also be selected by their IRAC colours based on the \cite{Donley12} criteria.
Of the 8 sources selected, 7 have good optical spectra (\citealt{Danielson17}) and/or 
X-ray photometry (Wang et al. 2013), and 3 of these show AGN signatures in the optical or X-ray.
Overall, 7 out of the 8 sources show a significant AGN signature from additional data 
(including SED fitting to multi-wavelength photometry). The remaining 1 source with none of 
the above mentioned signatures has a spectroscopic redshift of tentative quality (\citealt{Danielson17}), but 
does show a radio excess at 1.4GHz (based on flux density measurements in \citealt{Swinbank14}).
It is not surprising that we find SMGs hosting AGN, as mentioned in section~2.4, it is not uncommon for SMGs to 
exhibit AGN signatures. In addition to the sources discussed above, there are 7 sources that have been classified as hosts
of X-ray AGN \citep{Wang13} that are not selected by the F$_{870\um}$/F$_{24\um}$ selection limit, with 5 of them lying in the 
composite region of the F$_{870\um}$/F$_{24\um}$--$z$ plane (but above the selection limit), and 2 lying on the star formation 
region. The range of F$_{870\um}$/F$_{24\um}$ ratios of the SMGs with identified X-ray AGN, is not surprising given the range we
have already observed for the main sample of X-ray AGN in this work, and the moderate X-ray luminosities displayed by these 
sources (0.5--8keV luminosities of 10$^{42}$--10$^{44.5}\ergss$).

Overall, the F$_{870\um}$/F$_{24\um}$ selection limit based on Eqn.~1 
can successfully identify sources with strong AGN components in a variety of different samples. This 
demonstrates the potential of the F$_{870\um}$/F$_{24\um}$-redshift plane as a selection tool 
for AGN candidates, especially in the future where deep MIR, and sub-mm surveys will be available through 
observatories such as the James Webb Space Telescope (JWST) and ALMA.

\section{Summary \& Conclusions} \label{conclusion}
We use deep 870$\um$ ALMA observations to place constraints on the SFRs 
for a sample of 109 X-ray AGN that are faint or undetected in the {\it Herschel} bands.
Our sample covers X-ray luminosities of $10^{42} < \lx < 10^{45} \ergss$ at redshifts 
of $z =$1--4.7. 
Of our observed sample, 40/109 sources ($\sim$37\%) were detected at 870$\um$, 
but even though the majority are undetected
the flux limit provided by ALMA is sufficiently low to still place stronger 
constraints on the SFR limit value than previously possible.
We make use of the SED fitting methods of Stanley et al. (2015)
in combination with photometry at 8--870$\um$ to fit and decompose the IR SED into 
AGN and star-forming components. 

In summary: \begin{itemize}	
	\item We find that with the depths of our ALMA observations 40/109 ($\sim$37\%) of 
	our observed sample now have a measured SFR, 20 times more sources than 
	previously possible for this sample with 8--500$\um$ {\it Spitzer} and {\it Herschel} photometry. 
	Furthermore the majority of our sample, 73/109 ($\sim$67\%),
	now have SFR constraints that are factors of 2--10 lower than previously possible.
		
	\item With the excellent constraints at 870$\um$ on the star-forming component of the IR 
	SED, we are now able to place stronger constraints on the IR emission of the AGN.
	Indeed, we can now identify an AGN component in 54/109 ($\sim$50\%) of our ALMA observed sample, 
	with AGN fractions down to $\sim$20\% of the total IR emission, where without the ALMA photometry 
	we could identify a MIR AGN component in only one of the sources.
	
	\item We explore the parameter space of the flux density ratio of F$_{870\um}$/F$_{24\um}$
	with redshift, and find that it can clearly identify the presence of 
	MIR emission from the AGN, when the AGN contributes $\geq$50\% of the 
	total infrared emission. We test the F$_{870\um}$/F$_{24\um}$--redshift
	plane on two different comparison samples representing the two extremes of AGN and star formation
	dominated IR emission. We suggest that this method could be developed as a tool for identifying AGN 
	in future deep sub-mm and mid-infrared surveys (e.g., combining ALMA and JWST data). 
		
\end{itemize} 
Overall, we have demonstrated the importance of deep ALMA sub-mm observations for constraining the 
moderate to low SFRs of galaxies hosting AGN. With the build-up of deep ALMA observations of large galaxy 
samples we will be able to use the sub-mm to MIR colours, such as the F$_{870\um}$/F$_{24\um}$ ratio
to identify the presence of AGN emission in the IR. 

\subsection*{ACKNOWLEDGMENTS}

We thank the anonymous referee for their helpful comments on the improvement of this paper.
We acknowledge the Faculty of Science Durham Doctoral Scholarship (FS),
the Science and Technology Facilities Council (DMA,
DJR, through grant code ST/L00075X/1, JS through grant ST/N50404X/1), and the 
Leverhulme Trust (DMA). KK acknowledges support from the Knut and Alice Wallenberg Foundation. 
This paper makes use of ALMA data: ADS/JAO.ALMA\#2012.1.00869.S and ADS/JAO.ALMA\#2013.1.00884.S. 
ALMA is a partnership of ESO (representing its member states), NSF (USA) and NINS (Japan), 
together with NRC (Canada) and NSC and ASIAA (Taiwan), in cooperation with the Republic of Chile. 
The Joint ALMA Observatory is operated by ESO, AUI/NRAO and NAOJ.

\bibliography{full.bib}
\bibliographystyle{mn2e} 

\appendix
\renewcommand\thefigure{\thesection.\arabic{figure}}
\section{Source Tables and SED fits for our X-ray AGN sample and the two comparison samples}
In this Appendix section we present the best-fit SEDs and tabulated results, for our sample of X-ray AGN, and the 
two comparison samples of {\it WISE} AGN dominated sources and star forming galaxies from the ALESS survey (see section~5.1 for details).
Tables~A.1~\&~A.2 contain the source properties and best-fit SED results of our sample of X-ray AGN 
split into the two deep field of GOODS-S and C-COSMOS, while Tables~A.3~\&~A.4 contain the properties 
and best-fit SED results of the two comparison samples.
Fig.~A.1 contains the best-fit SEDs of our X-ray AGN sample, Fig.~A.2 the best-fit SEDs of the {\it WISE}
AGN dominated sample, and Fig.~A.3 the best-fit SEDs of the star forming galaxy sample.

\begin{table*}
	\begin{center}
		\begin{tabular}{|c|c|c|c|c|c|c|}
			\hline
			 Field & XID$^{(a)}$ & $z^{(b)}$ & L$_{\rm 2-8keV}$$^{(c)}$ & L$_{\rm IR,SF}$$^{(d)}$ & L$_{\rm IR,AGN}$$^{(e)}$  & AGN flag$^{(f)}$\\
			 			 	&		&    &    (erg/s)	&	($ \times 10^{45}$ erg/s)	   &  ($ \times 10^{45}$ erg/s)	&  \\
			\hline
			\hline
GS &  509 & 1.101 &  1.36$ \times 10^{ 42}$ & $<$0.66  & $<$0.1 & -1 \\
GS &  195 & 1.452 &  1.48$ \times 10^{ 43}$  &  1.09$^{+ 0.50}_{- 0.69}$ &  -- & 0 \\
GS &  167 & 1.455 &  2.12$ \times 10^{ 43}$  & $<$1.06  & 0.61$^{+ 0.09}_{- 0.09}$ & 2 \\
GS &  276 & 1.519 &  2.22$ \times 10^{ 42}$ &  1.38$^{+ 0.00}_{- 0.00}$ &  -- & 0 \\
GS &  257 & 1.536 &  0.85$ \times 10^{ 44}$  & $<$2.06  & 0.72$^{+ 0.16}_{- 0.16}$ & 2 \\
GS &  211 & 1.601 &  2.76$ \times 10^{ 43}$  & $<$1.24 & $<$0.32 & -1\\
GS &  184 & 1.605 &  3.11$ \times 10^{ 42}$  & $<$1.81  & $<$0.32 & -1\\
GS &  163 & 1.607 &  2.54$ \times 10^{ 42}$  & $<$2.04  & $<$0.51& -1\\
GS &  318 & 1.607 &  0.85$ \times 10^{ 42}$ &  1.64$^{+ 0.16}_{- 0.16}$ &  -- & 0\\
GS &  405 & 1.609 &  0.81$ \times 10^{ 43}$ & $<$1.08  & 0.65$^{+ 0.11}_{- 0.11}$ & 2\\
GS &  503 & 1.609 &  0.32$ \times 10^{ 43}$ & $<$2.06  & $<$0.32 & -1\\
GS &   88 & 1.616 &  0.55$ \times 10^{ 44}$ &  2.10$^{+ 0.23}_{- 0.23} $ &  0.90$^{+ 0.22}_{- 0.22} $ & 2\\
GS &  344 & 1.617 &  0.42$ \times 10^{ 44}$ &  8.49$^{+ 0.24}_{- 0.24} $ &  2.28$^{+ 0.36}_{- 0.36} $ & 2\\
GS &  308 & 1.727 &  0.72$ \times 10^{ 44}$ & $<$0.93  & 1.01$^{+ 0.15}_{- 0.15}$ & 2\\
GS &  221 & 1.887 &  0.50$ \times 10^{ 43}$  & $<$1.96  & $<$0.52 & -1\\
GS &  463 & 1.910 &  0.95$ \times 10^{ 42}$ & $<$2.27  & $<$0.54 & -1\\
GS &  155 & 2.019 &  2.05$ \times 10^{ 42}$  & $<$2.09  & $<$0.64 & -1\\
GS &  158 & 2.046 &  1.03$ \times 10^{ 43}$  & $<$2.78  & $<$0.66 & -1\\
GS &  522 & 2.115 &  0.36$ \times 10^{ 44}$ & $<$2.56  & 1.46$^{+ 0.31}_{- 0.31}$ & 2\\
GS &  388 & 2.129 &  0.88$ \times 10^{ 43}$ & $<$3.03  & $<$0.44 & -1\\
GS &  320 & 2.145 &  1.18$ \times 10^{ 43}$ &  2.70$^{+ 0.63}_{- 0.47}$ &  3.7$^{+ 0.4}_{- 0.6}$ & 2\\
GS &  277 & 2.209 &  0.50$ \times 10^{ 44}$ & $<$1.97  & $<$0.83 & -1\\
GS &  326 & 2.298 &  1.68$ \times 10^{ 43}$ & $<$1.73  &  0.73$^{+ 0.26}_{- 0.26}$ & 2\\
GS &  633 & 2.299 &  0.49$ \times 10^{ 44}$ &  2.28$^{+ 1.22}_{- 1.36}$ &  1.99$^{+ 0.77}_{- 0.75}$ & 2\\
GS &  123 & 2.331 &  1.01$ \times 10^{ 43}$  & $<$2.09  & $<$0.98 & -1\\
GS &  185 & 2.337 &  1.58$ \times 10^{ 42}$   &  0.89$^{+ 0.50}_{- 0.57}$ &  0.42$^{+ 0.31}_{- 0.28}$ & 2\\
GS &  310 & 2.392 &  2.65$ \times 10^{ 43}$ &  4.40$^{+ 0.37}_{- 0.37}$ &  -- & 0\\
GS &  444 & 2.393 &  0.50$ \times 10^{ 44}$ & $<$2.75  & 0.75$ \times 10^{45}$ & 2\\
GS &  215 & 2.402 &  2.50$ \times 10^{ 43}$  &  1.93$^{+ 1.06}_{- 1.23}$ &  1.43$^{+ 0.67}_{- 0.62}$ & 2\\
GS &  677 & 2.414 &  0.45$ \times 10^{ 44}$  & $<$4.72  & $<$1.09 & -1\\
GS &  199 & 2.417 &  2.33$ \times 10^{ 43}$  & $<$4.70  & $<$1.09 & -1\\
GS &  305 & 2.419 &  0.93$ \times 10^{ 43}$ & $<$2.16  & $<$1.09 & -1\\
GS &  574 & 2.427 &  0.87$ \times 10^{ 43}$  & $<$4.23  & $<$0.97 & -1\\
GS &  301 & 2.469 &  0.35$ \times 10^{ 44}$ &  7.51$^{+ 0.28}_{- 0.28}$ &  -- & 0 \\
GS &  422 & 2.492 &  0.50$ \times 10^{ 43}$ &  2.98$^{+ 0.71}_{- 2.09} $ &  -- & 0 \\
GS &  410 & 2.527 &  0.40$ \times 10^{ 44}$ &  2.78$^{+ 0.75}_{- 1.95} $ &  -- & 0 \\
GS &  351 & 2.532 &  2.50$ \times 10^{ 44}$ &  1.37$^{+ 0.24}_{- 0.24} $ &  7.38$^{+ 0.75}_{- 0.75}$ & 2\\
GS &  290 & 2.545 &  0.83$ \times 10^{ 44}$ & $<$6.56 & $<$1.55 & -1\\
GS &   93 & 2.573 &  0.69$ \times 10^{ 44}$  & $<$6.29  & $<$1.32 & -1\\
GS &  593 & 2.593 &  0.53$ \times 10^{ 44}$  & $<$6.44  & $<$1.54 & -1\\
GS &  137 & 2.610 &  1.66$ \times 10^{ 44}$  & $<$3.21  & 4.77$^{+ 0.56}_{- 0.56}$ & 2\\
GS &  294 & 2.611 &  0.39$ \times 10^{ 44}$ & $<$5.46  & $<$0.96 & -1\\
GS &  359 & 2.728 &  0.50$ \times 10^{ 44}$ & $<$6.56  & $<$1.57 & -1\\
GS &  466 & 2.775 &  2.99$ \times 10^{ 43}$ & $<$6.35 & 2.28$^{+ 0.68}_{- 0.68}$ & 2 \\
GS &  254 & 2.801 &  2.49$ \times 10^{ 43}$  & $<$6.01  & $<$1.70 & -1\\
GS &  528 & 2.973 &  0.56$ \times 10^{ 44}$ & $<$8.68 & 2.04$^{+ 0.65}_{- 0.65}$ & 2 \\
GS &  456 & 3.173 &  2.65$ \times 10^{ 43}$ & $<$1.93  & $<$2.47 & -1 \\
GS &  371 & 3.242 &  0.63$ \times 10^{ 44}$ &  7.60$^{+ 3.60}_{- 4.43}$ &  -- & 0\\
GS &  386 & 3.256 &  0.94$ \times 10^{ 43}$ &  1.58$^{+ 0.97}_{- 1.15}$ &  -- & 0\\
GS &  129 & 3.446 &  1.48$ \times 10^{ 44}$  & $<$1.82  & 3.56$^{+ 1.10}_{- 1.10}$ & 2 \\
GS &  262 & 3.660 &  1.61$ \times 10^{ 44}$  &  3.25$^{+ 0.93}_{- 0.82}$ &  2.35$^{+ 0.77}_{- 0.69}$ & 2 \\
GS &  412 & 3.700 &  2.70$ \times 10^{ 44}$ & $<$2.91  & 7.15$^{+ 0.93}_{- 0.93}$ & 2\\
GS &  230 & 3.985 &  1.32$ \times 10^{ 44}$  &  1.31$^{+ 0.71}_{- 0.88}$ &  -- & 0\\
GS &  534 & 4.379 &  0.85$ \times 10^{ 44}$ &  1.58$^{+ 0.68}_{- 1.24}$ &  10.57$^{+ 2.15}_{- 2.11}$ & 2 \\
GS &  156 & 4.651 &  0.91$ \times 10^{ 44}$  &  7.06$^{+ 2.37}_{- 3.69}$ &  -- & 0\\
		\end{tabular}
		\caption{Properties of the ALMA observed X-ray sample in the GOODS-S field.(a) The X-ray ID of the source in the \protect\cite{Xue11} catalogue; (b) the redshift of the source from \protect\cite{Hsu14}; (c) the X-ray hard-band luminosity of the source; (d) the IR luminosity due to star formation derived by the best fitting SED solution; (e) the IR luminosity due to the AGN derived from the the best fitting SED solution; (f) flag for the AGN component of the fits, where -1 = only upper limit constraints, 0 = SED fit does not require an AGN component, 1 = SED fit requires an AGN component, but has a weak contribution and is uncertain ($<$20\% of the IR luminosity), 2 = SED fit requires an AGN component with significant contribution ($\geq$20\% of the IR luminosity).} \label{tabresults_gs}
	\end{center}
\end{table*}

\begin{table*}
	\begin{center}
		\begin{tabular}{|c|c|c|c|c|c|c|}
			\hline
			 Field & XID$^{(a)}$ & $z^{(b)}$ & L$_{\rm 2-8keV}$$^{(c)}$ & L$_{\rm IR,SF}$$^{(d)}$ & L$_{\rm IR,AGN}$$^{(e)}$ & AGN flag$^{(f)}$ \\
			 	&		&    &    (erg/s)	&	($ \times 10^{45}$ erg/s)	   &  ($ \times 10^{45}$ erg/s)	& \\
			\hline
			\hline
 C &   85 & 1.349 &  0.89$ \times 10^{ 44}$  & $<$1.10  & $<$0.19 & -1\\
 C &  434 & 1.530 &  0.51$ \times 10^{ 45}$ &  0.79$^{+ 0.52}_{- 0.53}$ & 1.84$^{+ 0.47}_{- 0.47}$ & 2\\
 C & 1214 & 1.594 &  1.41$ \times 10^{ 44}$ &  0.88$^{+ 0.63}_{- 0.65}$ &  -- & 0 \\
 C &   87 & 1.598 &  1.01$ \times 10^{ 45}$ &  0.63$^{+ 0.50}_{- 0.50} $ &  3.39$^{+ 0.60}_{- 0.60}$ & 2\\
 C &  581 & 1.708 &  0.39$ \times 10^{ 45}$  & $<$1.08  & $<$0.39 & -1\\
 C &  330 & 1.753 &  0.57$ \times 10^{ 45}$ &  0.61$^{+ 0.38}_{- 0.41}$ &  5.33$^{+ 1.25}_{- 1.25}$ & 2\\
 C &   53 & 1.787 &  2.22$ \times 10^{ 44}$  & $<$1.69  & $<$0.44 & -1\\
 C &  474 & 1.796 &  0.41$ \times 10^{ 45}$  & $<$1.15  & 5.21$^{+ 0.95}_{- 0.95}$ & 2 \\
 C &  532 & 1.796 &  0.38$ \times 10^{ 45}$  & $<$1.11 & 2.50$^{+ 0.60}_{- 0.60}$ & 2 \\
 C &   86 & 1.831 &  2.87$ \times 10^{ 44}$ &  0.59$^{+ 0.45}_{- 0.47}$ &  -- & 0 \\
 C &  915 & 1.841 &  1.37$ \times 10^{ 44}$  & $<$1.19  & 4.37$^{+ 1.04}_{- 1.00}$ & 2\\
 C &  987 & 1.860 &  1.33$ \times 10^{ 44}$  & $<$1.68  & 0.67$^{+ 0.25}_{- 0.25}$ & 2 \\
 C & 1144 & 1.912 &  1.61$ \times 10^{ 44}$ &  1.48$^{+ 1.17}_{- 0.96}$ &  -- & 0 \\
 C &   62 & 1.914 &  0.40$ \times 10^{ 45}$ &  1.37$^{+ 0.84}_{- 0.90}$ &  0.34$^{+ 0.09}_{- 0.09} \times 10^{46}$ & 2\\
 C &   90 & 1.932 &  0.38$ \times 10^{ 45}$  & $<$1.25  & 1.36$^{+ 0.42}_{- 0.42}$ & 2\\
 C &  954 & 1.936 &  2.32$ \times 10^{ 44}$ &  1.07$^{+ 0.65}_{- 0.70} \times 10^{45}$ &  0.36$^{+ 0.10}_{- 0.09} \times 10^{46}$ & 2\\
 C &   81 & 1.991 &  1.55$ \times 10^{ 44}$ &  0.87$^{+ 0.52}_{- 0.57} \times 10^{45}$ &  2.80$^{+ 0.79}_{- 0.78} \times 10^{45}$ & 2\\
 C &  351 & 2.018 &  0.57$ \times 10^{ 45}$  & $<$1.69  & 3.36$^{+ 0.88}_{- 0.88}$ & 2\\
 C &  659 & 2.045 &  1.50$ \times 10^{ 44}$  & $<$1.16  & 1.74$^{+ 0.52}_{- 0.52}$ & 2\\
 C &  580 & 2.111 &  0.43$ \times 10^{ 45}$  & $<$1.73   & 0.96$^{+ 0.38}_{- 0.38}$ & 2\\
 C &  706 & 2.113 &  1.16$ \times 10^{ 44}$  & $<$1.74  & $<$0.73 & -1 \\
 C &  960 & 2.122 &  1.06$ \times 10^{ 44}$ &  1.31$^{+ 0.76}_{- 0.85}$ &  2.15$^{+ 0.74}_{- 0.72}$ & 2\\
 C &  914 & 2.146 &  1.57$ \times 10^{ 44}$ &  1.36$^{+ 0.79}_{- 0.89}$ &  0.65$^{+ 0.50}_{- 0.47}$ & 2\\
 C & 1620 & 2.169 &  0.53$ \times 10^{ 45}$  & $<$1.66  & 3.77$^{+ 0.97}_{- 0.97}$ & 2\\
 C & 1085 & 2.231 &  0.52$ \times 10^{ 45}$  & $<$1.82  & 2.69$^{+ 0.76}_{- 0.76}$ & 2\\
 C & 1205 & 2.255 &  1.22$ \times 10^{ 44}$ &  0.96$^{+ 0.68}_{- 0.74}$ &  -- & 0 \\
 C &  467 & 2.288 &  0.94$ \times 10^{ 45}$ &  1.05$^{+ 0.59}_{- 0.68}$ &  8.51$^{+ 2.02}_{- 2.01}$ & 2\\
 C & 1127 & 2.390 &  1.99$ \times 10^{ 44}$  & $<$2.70  & 1.29$^{+ 0.49}_{- 0.49}$ & 2\\
 C &  451 & 2.450 &  0.65$ \times 10^{ 45}$ &  0.67$^{+ 0.56}_{- 0.60} $ &  27.34$^{+ 3.86}_{- 3.86}$ & 2\\
 C & 1215 & 2.450 &  1.85$ \times 10^{ 44}$ &  6.37$^{+ 0.64}_{- 2.67}$ &  5.72$^{+ 1.28}_{- 1.77}$ & 2\\
 C & 1143 & 2.454 &  1.54$ \times 10^{ 44}$  & $<$1.68  & $<$1.14 & -1\\
 C &   72 & 2.475 &  0.56$ \times 10^{ 45}$  & $<$1.80  & 4.11$^{+ 1.25}_{- 1.25}$ & 2\\
 C &  976 & 2.478 &  1.14$ \times 10^{ 44}$  & $<$1.63 & $<$1.18 & -1\\
 C &  352 & 2.498 &  0.63$ \times 10^{ 45}$  & $<$1.67  & 12.73$^{+ 2.92}_{- 2.92}$ & 2\\
 C &  970 & 2.501 &  0.64$ \times 10^{ 45}$  & $<$2.55  & 6.16$^{+ 1.73}_{- 1.73}$ & 2\\
 C &  708 & 2.548 &  1.42$ \times 10^{ 44}$ &  1.93$^{+ 1.03}_{- 1.22}$ &  1.00$^{+ 0.68}_{- 0.63}$ & 2 \\
 C &   31 & 2.611 &  0.90$ \times 10^{ 45}$ &  1.77$^{+ 0.93}_{- 1.12}$ &  8.1$^{+ 3.1}_{- 3.1}$ & 2 \\
 C & 1216 & 2.663 &  1.84$ \times 10^{ 44}$  & $<$2.68 & 4.92$^{+ 1.89}_{- 1.82}$ & 2 \\
 C &  365 & 2.671 &  0.55$ \times 10^{ 45}$  & $<$2.59  & 11.60$^{+ 2.75}_{- 2.75}$ & 2 \\
 C &  121 & 2.791 &  0.43$ \times 10^{ 45}$  & $<$2.72  & $<$1.68 & -1 \\
 C &   58 & 2.798 &  0.56$ \times 10^{ 45}$  & $<$2.60  & $<$1.69 & -1 \\
 C &  459 & 2.887 &  0.86$ \times 10^{ 45}$  & $<$2.72  & 19.41$^{+ 3.28}_{- 3.28}$ & 2\\
 C & 1246 & 2.888 &  1.75$ \times 10^{ 44}$  & $<$2.70  & $<$1.86 & -1 \\
 C & 1219 & 2.946 &  2.23$ \times 10^{ 44}$  & $<$2.53 & $<$1.98 & -1 \\
 C &  149 & 2.955 &  0.62$ \times 10^{ 45}$  & $<$2.91  & 2.78$^{+ 1.04}_{- 1.04}$ & 2 \\
 C &  529 & 3.017 &  0.61$ \times 10^{ 45}$  & $<$2.82  & $<$2.12 & -1 \\
 C &   75 & 3.029 &  0.86$ \times 10^{ 45}$ &  0.98$^{+ 0.83}_{- 0.91}$ &  27.58$^{+ 5.07}_{- 5.07}$ & 2 \\
 C &  124 & 3.072 &  0.37$ \times 10^{ 45}$  & $<$2.65  & $<$2.24 & -1\\
 C &   83 & 3.075 &  0.55$ \times 10^{ 45}$  & $<$2.55 & $<$2.25 & -1\\
 C & 1247 & 3.087 &  1.21$ \times 10^{ 44}$  & $<$2.54  & $<$2.27 & -1\\
 C &  917 & 3.090 &  1.45$ \times 10^{ 44}$ &  9.24$^{+ 4.27}_{- 5.26} $ &  -- & 0 \\
 C &  953 & 3.095 &  1.98$ \times 10^{ 44}$  & $<$2.72  & $<$2.29 & -1\\
 C &  558 & 3.100 &  0.95$ \times 10^{ 45}$  & $<$2.70  & $<$2.30 & -1\\
 C &  965 & 3.178 &  2.86$ \times 10^{ 44}$ &  1.79$^{+ 0.83}_{- 1.08}$ &  11.62$^{+ 3.00}_{- 3.00}$ & 2\\
		\end{tabular}
		\caption{Properties of the ALMA observed X-ray sample in the COMSOS field. (a) The X-ray ID of the source in the \protect\cite{Civano12} catalogue; (b) the redshift of the source from \protect\cite{Marchesi16}; (c) the 2--10keV luminosity of the source; (d) the IR luminosity due to star formation derived by the best fitting SED solution; (e) the IR luminosity due to the AGN derived from the the best fitting SED solution; (f) flag for the AGN component of the fits, where -1 = only upper limit constraints, 0 = SED fit does not require an AGN component, 1 = SED fit requires an AGN component, but has a weak contribution and is uncertain ($<$20\% of the IR luminosity), 2 = SED fit requires an AGN component with significant contribution ($\geq$20\% of the IR luminosity).} \label{tabresults_c}
	\end{center}
\end{table*}

\begin{table*}
	\begin{center}
		\begin{tabular}{|c|c|c|c|c|}
			\hline
			 WISE-ID$^{(a)}$ & z$^{(b)}$  & L$_{IR,SF}$$^{(c)}$ & L$_{IR,AGN}$$^{(d)}$ & AGN flag$^{(e)}$ \\
			  		&    &    ($\times 10^{46}$ erg/s)	&	($\times 10^{46}$ erg/s)	 & 	\\
			\hline
			\hline
 W1514-3411 & 1.090 & $<0.34$ & 1.72$^{+ 0.35}_{- 0.35}$  & 2 \\
 W0811-2225 & 1.110 & $<0.68$ & 2.21$^{+ 0.48}_{- 0.48}$  & 2 \\
 W1439-3725 & 1.190 & $<0.23$ & 1.41$^{+ 0.32}_{- 0.32}$  & 2 \\
 W0404-2436 & 1.260 &  0.85$^{+ 0.53}_{- 0.61}$ &  1.95$^{+ 0.50}_{- 0.50}$ & 2  \\
 W0642-2728 & 1.340 &  0.59$^{+ 0.44}_{- 0.43}$ &  1.41$^{+ 0.39}_{- 0.39}$  & 2 \\
 W0354-3308 & 1.370 & $<$0.49 & 3.40$^{+ 0.65}_{- 0.65}$  & 2 \\
 W0630-2120 & 1.440 &  1.37$^{+ 1.02}_{- 1.00}$ &  2.80$^{+ 0.66}_{- 0.66}$  & 2 \\
 W1500-0649 & 1.500 &  2.99$^{+ 0.13}_{- 0.13}$ &  9.64$^{+ 0.39}_{- 0.39} $ & 2 \\
 W0304-3108 & 1.540 &  0.69$^{+ 0.47}_{- 0.48}$ &  6.77$^{+ 1.19}_{- 1.19} $ & 2  \\
 W1541-1144 & 1.580 &  0.26$^{+ 0.19}_{- 0.20}$ &  5.78$^{+ 1.15}_{- 1.15} $  & 2 \\
 W1951-0420 & 1.580 & $<$0.44 & 5.02$^{+ 1.00}_{- 1.00}$  & 2 \\
 W0719-3349 & 1.630 &  1.33$^{+ 1.14}_{- 0.86}$ &  3.17$^{+ 0.77}_{- 0.80}$  & 2 \\
 W1308-3447 & 1.650 &  0.29$^{+ 0.22}_{- 0.22}$ &  6.61$^{+ 1.21}_{- 1.21}$  & 2 \\
 W1400-2919 & 1.670 & $<$0.38 & 9.92$^{+ 1.72}_{- 1.72}$ & 2 \\
 W0525-3614 & 1.690 & $<$0.63 & 3.11$^{+ 0.68}_{- 0.68}$ & 2  \\
 W0549-3739 & 1.710 &  0.55$^{+ 0.38}_{- 0.39}$ &  2.44$^{+ 0.60}_{- 0.60}$  & 2 \\
 W0823-0624 & 1.750 & $<$0.76 & 9.44$^{+ 1.72}_{- 1.72}$ & 2 \\
 W0536-2703 & 1.790 &  0.70$^{+ 0.46}_{- 0.48}$ &  5.97$^{+ 1.27}_{- 1.27}$  & 2 \\
 W1703-0517 & 1.800 &  0.18$^{+ 0.14}_{- 0.15}$ &  7.14$^{+ 1.63}_{- 1.63}$ & 2  \\
 W1958-0746 & 1.800 & $<$0.39 & 7.79$^{+ 1.56}_{- 1.56}$  & 2 \\
 W1412-2020 & 1.820 &  0.62$^{+ 0.44}_{- 0.46} $ &  7.74$^{+ 1.50}_{- 1.50}$  & 2 \\
 W1641-0548 & 1.840 &  0.57$^{+ 0.37}_{- 0.39} $ &  7.06$^{+ 1.48}_{- 1.48}$ & 2  \\
 W1434-0235 & 1.920 & $<$0.38 &  6.19$^{+ 1.24}_{- 1.24}$ & 2 \\
 W0526-3225 & 1.980 &  5.04$^{+ 3.10}_{- 3.33} $ &  24.28$^{+ 4.25}_{- 4.24}$  & 2 \\
 W0614-0936 & 2.000 & $<$0.77 & 7.52$^{+ 1.55}_{- 1.55}$  & 2 \\
 W1657-1740 & 2.000 & $<$0.33 & 10.07$^{+ 2.11}_{- 2.11}$ & 2 \\
 W1707-0939 & 2.000 & $<$0.43 &  6.39$^{+ 1.89}_{- 1.89}$ & 2 \\
 W2040-3904 & 2.000 &  2.30$^{+ 0.22}_{- 1.83}$ &  6.68$^{+ 1.56}_{- 1.61}$ & 2 \\
 W1653-0102 & 2.020 & $<$0.33 & 8.04$^{+ 1.69}_{- 1.69}$ & 2 \\
 W0519-0813 & 2.050 & $<$0.64 & 6.76$^{+ 1.45}_{- 1.45}$ & 2 \\
 W1634-1721 & 2.080 & $<$0.36 & 6.05$^{+ 1.55}_{- 1.55}$ & 2 \\
 W0613-3407 & 2.180 & $<$0.77 & 11.79$^{+ 2.18}_{- 2.18}$ & 2 \\
 W1513-2210 & 2.200 &  1.25$^{+ 0.74}_{- 0.82}$ &  12.38$^{+ 2.55}_{- 2.55}$ & 2 \\
 W1936-3354 & 2.240 &  0.38$^{+ 0.25}_{- 0.28}$ &  9.87$^{+ 2.15}_{- 2.15}$ & 2 \\
 W2000-2802 & 2.280 & $<$0.41 & 13.97$^{+ 2.99}_{- 2.99}$ & 2 \\
 W2059-3541 & 2.380 & $<$0.42 & 4.26$^{+ 0.79}_{- 0.79}$ & 2 \\
 W2021-2611 & 2.440 &  1.25$^{+ 0.71}_{- 0.81} $ &  6.69$^{+ 1.85}_{- 1.85}$ & 2 \\
 W1343-1136 & 2.490 &  0.54$^{+ 0.32}_{- 0.36}$ &  9.15$^{+ 2.11}_{- 2.11}$ & 2\\
 W1521+0017 & 2.630 &  -- &  22.12$^{+ 3.26}_{- 3.26}$ & 2 \\
 W0439-3159 & 2.820 &  1.69$^{+ 0.89}_{- 1.07}$ &  10.22$^{+ 2.37}_{- 2.37}$ & 2 \\
 W1702-0811 & 2.850 & $<$0.43 & 25.91$^{+ 6.85}_{- 6.85}$ & 2 \\
		\end{tabular}
		\caption{SED fitting results for the comparison sample of {\it WISE} AGN dominated sources. (a) The {\it WISE} ID of the source as given in \protect\cite{Lonsdale15}; (b) the redshift of the source from \protect\cite{Lonsdale15}; (c) the IR luminosity due to star formation derived by the best fitting SED solution; (d) the IR luminosity due to the AGN derived from the the best fitting SED solution; ; (e) flag for the AGN component of the fits, where -1 = only upper limit constraints, 0 = SED fit does not require an AGN component, 1 = SED fit requires an AGN component, but has a weak contribution and is uncertain ($<$20\% of the IR luminosity), 2 = SED fit requires an AGN component with significant contribution ($\geq$20\% of the IR luminosity).} \label{tab_hd_results}
	\end{center}
\end{table*}

\begin{table*}
	\begin{center}
		\begin{tabular}{|c|c|c|c|c|}
			\hline
			 ALESS-ID$^{(a)}$ & z$^{(b)}$ & L$_{IR,SF}$$^{(c)}$ & L$_{IR,AGN}$$^{(d)}$ & AGN flag$^{(e)}$ \\
			  		&    &    ($\times 10^{46}$ erg/s)	&	($\times 10^{45}$ erg/s)	 & 	\\
			\hline
			\hline
ALESS 103.2 & 1.000 &  0.63$^{+ 0.08}_{- 0.60}$ &  --  & 0 \\
ALESS 089.1 & 1.170 &  0.41$^{+ 0.07}_{- 0.34}$ &  0.12$^{+ 0.03}_{- 0.12}$ & 1 \\
ALESS 088.1 & 1.268 &  0.55$^{+ 0.32}_{- 0.38}$ &  -- & 0 \\
ALESS 062.2 & 1.361 &  1.06$^{+ 0.05}_{- 0.84} $ &  -- & 0 \\
ALESS 051.1 & 1.363 &  0.21$^{+ 0.23}_{- 0.02}$ &  -- & 0 \\
ALESS 080.2 & 1.365 &  0.34$^{+ 0.04}_{- 0.18}$ &  -- & 0 \\
ALESS 098.1 & 1.373 &  0.91$^{+ 0.06}_{- 0.67}$ &  -- & 0 \\
ALESS 055.1 & 1.378 &  0.17$^{+ 0.09}_{- 0.00} $ &  -- & 0 \\
ALESS 003.2 & 1.390 &  0.16$^{+ 0.18}_{- 0.03} $ &  -- & 0 \\
ALESS 029.1 & 1.439 &  1.61$^{+ 0.11}_{- 1.40} $ &  -- & 0 \\
ALESS 049.2 & 1.465 &  0.44$^{+ 0.05}_{- 0.28} $ &  -- & 0 \\
ALESS 084.2 & 1.471 &  0.49$^{+ 0.11}_{- 0.32} $ &  0.08$^{+ 0.40}_{- 0.16}$ & 1 \\
ALESS 063.1 & 1.490 &  0.73$^{+ 0.00}_{- 0.49} $ &  --  & 0 \\
ALESS 017.1 & 1.540 &  0.95$^{+ 0.05}_{- 0.62} $ &  --  & 0 \\
ALESS 114.2 & 1.606 &  0.97$^{+ 0.07}_{- 0.75} $ &  0.32$^{+ 0.14}_{- 0.35}$ & 1  \\
ALESS 043.1 & 1.705 &  0.38$^{+ 0.45}_{- 0.12} $ &  --  & 0 \\
ALESS 079.2 & 1.769 &  0.81$^{+ 0.19}_{- 0.56} $ &  1.90$^{+ 2.37}_{- 2.02}$ & 2 \\
ALESS 074.1 & 1.800 &  0.86$^{+ 0.80}_{- 0.58} $ &  --  & 0 \\
ALESS 126.1 & 1.815 &  0.54$^{+ 0.08}_{- 0.30} $ &  --  & 0 \\
ALESS 092.2 & 1.900 &  0.28$^{+ 0.11}_{- 0.14} $ &  --  & 0 \\
ALESS 015.1 & 1.925 &  1.72$^{+ 1.21}_{- 1.06} $ &  --  & 0 \\
ALESS 043.3 & 1.975 &  0.29$^{+ 0.22}_{- 0.13} $ &  --  & 0 \\
ALESS 122.1 & 2.025 &  2.19$^{+ 0.15}_{- 1.75} $ &  9.24$^{+ 17.91}_{- 2.42}$ & 2 \\
ALESS 079.1 & 2.045 &  1.50$^{+ 0.00}_{- 0.00} $ &  -- & 0 \\
ALESS 059.2 & 2.090 &  0.93$^{+ 0.14}_{- 0.71} $ &  -- & 0 \\
ALESS 070.1 & 2.093 &  2.58$^{+ 0.19}_{- 1.99} $ &  4.12$^{+ 1.07}_{- 4.25}$ & 1 \\
ALESS 082.1 & 2.095 &  2.06$^{+ 0.00}_{- 0.00} $ &  -- & 0 \\
ALESS 075.4 & 2.100 &  2.88$^{+ 2.68}_{- 1.43} $ &  -- & 0 \\
ALESS 107.3 & 2.115 &  2.96$^{+ 3.30}_{- 1.90} $ &  -- & 0 \\
ALESS 067.1 & 2.135 &  1.49$^{+ 0.40}_{- 0.94} $ &  -- & 0 \\
ALESS 081.1 & 2.145 &  2.40$^{+ 0.21}_{- 1.81} $ &  3.99$^{+ 3.79}_{- 4.18}$ & 1 \\
ALESS 019.2 & 2.170 &  0.34$^{+ 0.50}_{- 0.25} $ &  -- & 0 \\
ALESS 002.1 & 2.191 &  1.68$^{+ 0.13}_{- 0.13} $ &  -- & 0 \\
ALESS 022.1 & 2.266 &  2.26$^{+ 0.21}_{- 1.73} $ &  6.17$^{+ 1.45}_{- 5.93}$ & 2 \\
ALESS 088.5 & 2.291 &  1.57$^{+ 0.16}_{- 1.21} $ &  1.60$^{+ 0.47}_{- 1.67}$ & 1 \\
ALESS 075.2 & 2.294 &  0.56$^{+ 0.20}_{- 0.18} $ &  -- & 0 \\
ALESS 102.1 & 2.296 &  0.82$^{+ 0.43}_{- 0.44} $ &  -- & 0 \\
ALESS 112.1 & 2.314 &  1.78$^{+ 1.26}_{- 1.12} $ &  -- & 0 \\
ALESS 087.1 & 2.318 &  0.80$^{+ 0.15}_{- 0.65} $ &  8.20$^{+ 1.68}_{- 2.48} $ & 2 \\
ALESS 006.1 & 2.330 &  1.30$^{+ 0.93}_{- 0.89} $ &  -- & 0 \\
ALESS 045.1 & 2.340 &  2.24$^{+ 0.17}_{- 1.98} $ &  -- & 0 \\
ALESS 055.5 & 2.345 &  0.35$^{+ 0.25}_{- 0.23} $ &  -- & 0 \\
ALESS 093.1 & 2.350 &  0.89$^{+ 0.76}_{- 0.52} $ &  -- & 0 \\
ALESS 083.1 & 2.360 &  1.80$^{+ 0.21}_{- 0.21} $ &  -- & 0 \\
ALESS 062.1 & 2.380 &  0.87$^{+ 0.95}_{- 0.60} $ &  -- & 0 \\
ALESS 019.1 & 2.410 &  2.32$^{+ 0.17}_{- 2.03} $ &  -- & 0 \\
ALESS 118.1 & 2.413 &  1.53$^{+ 0.19}_{- 1.24} $ &  -- & 0 \\
ALESS 039.1 & 2.440 &  1.13$^{+ 0.71}_{- 0.68} $ &  -- & 0 \\
ALESS 017.2 & 2.441 &  0.55$^{+ 0.22}_{- 0.22} $ &  -- & 0 \\
ALESS 037.2 & 2.463 &  0.42$^{+ 0.31}_{- 0.29} $ &  -- & 0 \\
ALESS 038.1 & 2.470 &  1.60$^{+ 0.17}_{- 0.99} $ &  -- & 0 \\
ALESS 034.1 & 2.510 &  1.01$^{+ 0.89}_{- 0.59} $ &  -- & 0 \\
ALESS 110.1 & 2.545 &  1.50$^{+ 0.00}_{- 1.05} $ &  -- & 0 \\
ALESS 041.1 & 2.546 &  1.90$^{+ 0.38}_{- 1.25} $ &  -- & 0 \\
ALESS 066.1 & 2.554 &  1.24$^{+ 0.18}_{- 0.96} $ &  12.33$^{+ 3.15}_{- 2.92}$ & 2 \\
\end{tabular}
		\caption{SED fitting results for the comparison sample of star formation galaxies. (a) The ALESS ID of the source as given in \protect\cite{Hodge13}; (b) the redshift of the source from \protect\cite{Simpson14} and \protect\cite{Danielson17}; (c) the IR luminosity due to star formation derived by the best fitting SED solution; (d) the IR luminosity due to the AGN derived from the the best fitting SED solution; (e) flag for the AGN component of the fits, where -1 = only upper limit constraints, 0 = SED fit does not require an AGN component, 1 = SED fit requires an AGN component, but has a weak contribution and is uncertain ($<$20\% of the IR luminosity), 2 = SED fit requires an AGN component with significant contribution ($\geq$20\% of the IR luminosity).} \label{tab_smg_results}
	\end{center}
\end{table*}

\begin{table*}
	\begin{center}
		\begin{tabular}{|c|c|c|c|c|}
			\hline
			 ALESS-ID$^{(a)}$ & z$^{(b)}$ & L$_{IR,SF}$$^{(c)}$ & L$_{IR,AGN}$$^{(d)}$ & AGN flag$^{(e)}$\\
			  		&    &    ($\times 10^{46}$ erg/s)	&	($\times 10^{45}$ erg/s)	 & 	\\
			\hline
			\hline
ALESS 075.1 & 2.556 &  1.09$^{+ 0.50}_{- 0.72} $ &  17.40$^{+ 4.98}_{- 5.82}$ & 2 \\
ALESS 088.1 & 2.565 &  1.40$^{+ 0.18}_{- 0.99} $ &  -- & 0 \\
ALESS 017.3 & 2.575 &  0.76$^{+ 0.08}_{- 0.11} $ &  -- & 0 \\
ALESS 020.1 & 2.575 &  1.42$^{+ 0.27}_{- 0.79} $ &  --  & 0 \\
ALESS 011.1 & 2.680 &  1.93$^{+ 1.17}_{- 1.14} $ &  --  & 0 \\
ALESS 018.1 & 2.689 &  2.56$^{+ 0.21}_{- 1.99} $ &  4.83$^{+ 2.88}_{- 4.66} $ & 1 \\
ALESS 007.1 & 2.693 &  2.67$^{+ 0.23}_{- 1.95} $ &  6.30$^{+ 5.92}_{- 2.53} $ & 2 \\
ALESS 071.3 & 2.725 &  0.36$^{+ 0.25}_{- 0.24} $ &  -- & 0 \\
ALESS 049.1 & 2.760 &  2.34$^{+ 0.19}_{- 1.60} $ &  -- & 0 \\
ALESS 101.1 & 2.800 &  1.54$^{+ 0.22}_{- 1.26} $ &  -- & 0 \\
ALESS 001.3 & 2.845 &  1.09$^{+ 0.18}_{- 0.86} $ &  -- & 0 \\
ALESS 005.1 & 2.860 &  2.87$^{+ 0.24}_{- 0.24} $ &  -- & 0 \\
ALESS 094.1 & 2.870 &  1.18$^{+ 0.24}_{- 0.78} $ &  2.72$^{+ 2.39}_{- 1.13}$ & 1 \\
ALESS 025.1 & 2.880 &  3.23$^{+ 0.19}_{- 2.46} $ &  4.60$^{+ 1.05}_{- 4.58}$ & 1 \\
ALESS 031.1 & 2.885 &  3.44$^{+ 0.16}_{- 0.16} $ &  -- & 0 \\
ALESS 023.7 & 2.900 &  0.46$^{+ 0.31}_{- 0.31} $ &  -- & 0 \\
ALESS 057.1 & 2.938 &  1.85$^{+ 0.24}_{- 1.39} $ &  9.11$^{+ 1.93}_{- 3.18} $ & 2 \\
ALESS 114.1 & 3.000 &  1.91$^{+ 0.27}_{- 1.46} $ &  -- & 0 \\
ALESS 107.1 & 3.048 &  1.05$^{+ 0.17}_{- 0.81} $ &  -- & 0 \\
ALESS 001.2 & 3.086 &  0.91$^{+ 0.42}_{- 0.53} $ &  -- & 0 \\
ALESS 041.3 & 3.100 &  0.72$^{+ 0.47}_{- 0.47} $ &  -- & 0 \\
ALESS 013.1 & 3.250 &  2.04$^{+ 0.58}_{- 0.99} $ &  -- & 0 \\
ALESS 035.1 & 3.300 &  2.20$^{+ 0.15}_{- 1.63} $ &  -- & 0 \\
ALESS 030.1 & 3.360 &  1.29$^{+ 0.75}_{- 0.78} $ &  -- & 0 \\
ALESS 081.2 & 3.370 &  0.64$^{+ 0.39}_{- 0.40} $ &  -- & 0 \\
ALESS 076.1 & 3.390 &  1.28$^{+ 0.53}_{- 0.51} $ &  -- & 0 \\
ALESS 001.1 & 3.435 &  1.56$^{+ 0.57}_{- 0.76} $ &  -- & 0 \\
ALESS 015.3 & 3.441 &  0.53$^{+ 0.32}_{- 0.34} $ &  -- & 0 \\
ALESS 119.1 & 3.500 &  1.48$^{+ 0.33}_{- 0.48} $ &  -- & 0 \\
ALESS 037.1 & 3.530 &  1.52$^{+ 0.18}_{- 1.14} $ &  15.62$^{+ 3.07}_{- 3.09} $ & 2 \\
ALESS 116.1 & 3.540 &  1.68$^{+ 0.20}_{- 1.26} $ &  -- & 0 \\
ALESS 023.2 & 3.555 &  1.45$^{+ 0.79}_{- 0.84} $ &  -- & 0 \\
ALESS 072.1 & 3.596 &  1.36$^{+ 0.71}_{- 0.76} $ &  -- & 0 \\
ALESS 035.2 & 3.700 &  0.39$^{+ 0.23}_{- 0.24} $ &  -- & 0 \\
ALESS 110.5 & 3.700 &  0.66$^{+ 0.38}_{- 0.41} $ &  -- & 0 \\
ALESS 071.1 & 3.701 &  1.89$^{+ 0.27}_{- 1.59} $ &  89.63 $^{+ 8.92}_{- 6.42}$ & 2 \\
ALESS 115.0 & 3.789 &  3.36$^{+ 0.19}_{- 2.39} $ &  -- & 0 \\
ALESS 067.2 & 3.881 &  0.90$^{+ 0.19}_{- 0.68} $ &  -- & 0 \\
ALESS 002.2 & 3.920 &  2.25$^{+ 0.27}_{- 1.61} $ &  -- & 0 \\
ALESS 084.1 & 3.965 &  2.00$^{+ 0.27}_{- 1.55} $ &  20.26$^{+ 3.70}_{- 3.28} $ & 2 \\
ALESS 087.3 & 4.000 &  0.70$^{+ 0.38}_{- 0.41} $ &  -- & 0 \\
ALESS 116.2 & 4.015 &  1.89$^{+ 0.24}_{- 1.37} $ &  -- & 0 \\
ALESS 055.2 & 4.200 &  0.69$^{+ 0.36}_{- 0.40} $ &  -- & 0 \\
ALESS 069.1 & 4.211 &  1.42$^{+ 0.63}_{- 0.73} $ &  7.52$^{+ 1.94}_{- 1.83} $ & 2 \\
ALESS 003.1 & 4.237 &  4.00$^{+ 0.18}_{- 2.73} $ &  -- & 0 \\
ALESS 103.3 & 4.400 &  0.43$^{+ 0.22}_{- 0.25} $ &  -- & 0 \\
ALESS 061.1 & 4.440 &  2.05$^{+ 0.23}_{- 1.39} $ &  -- & 0 \\
ALESS 065.1 & 4.444 &  1.25$^{+ 0.54}_{- 0.64} $ &  -- & 0 \\
ALESS 014.1 & 4.465 &  3.75$^{+ 0.24}_{- 2.55} $ &  7.45$^{+ 1.62}_{- 3.03} $ & 1 \\
ALESS 009.1 & 4.500 &  4.21$^{+ 0.22}_{- 2.79} $ &  -- & 0 \\
ALESS 079.4 & 4.600 &  0.55$^{+ 0.28}_{- 0.32} $ &  -- & 0 \\
ALESS 080.1 & 4.660 &  1.24$^{+ 0.57}_{- 0.66} $ &  -- & 0 \\
ALESS 069.2 & 4.750 &  0.73$^{+ 0.34}_{- 0.40} $ &  -- & 0 \\
ALESS 073.1 & 4.755 &  1.77$^{+ 0.75}_{- 0.79} $ &  -- & 0 \\
ALESS 069.3 & 4.800 &  0.64$^{+ 0.31}_{- 0.36} $ &  -- & 0 \\
ALESS 023.1 & 4.990 &  3.29$^{+ 0.17}_{- 2.08}$ &  -- & 0 \\
ALESS 099.1 & 5.000 &  0.66$^{+ 0.28}_{- 0.33}$ & -- & 0 \\
		\end{tabular}
		\caption{SED fitting results for the comparison sample of star forming galaxies (continued).} 
	\end{center}
\end{table*}

\setcounter{figure}{0}
\begin{figure*}  
	\begin{center}
		\includegraphics[scale=1.2]{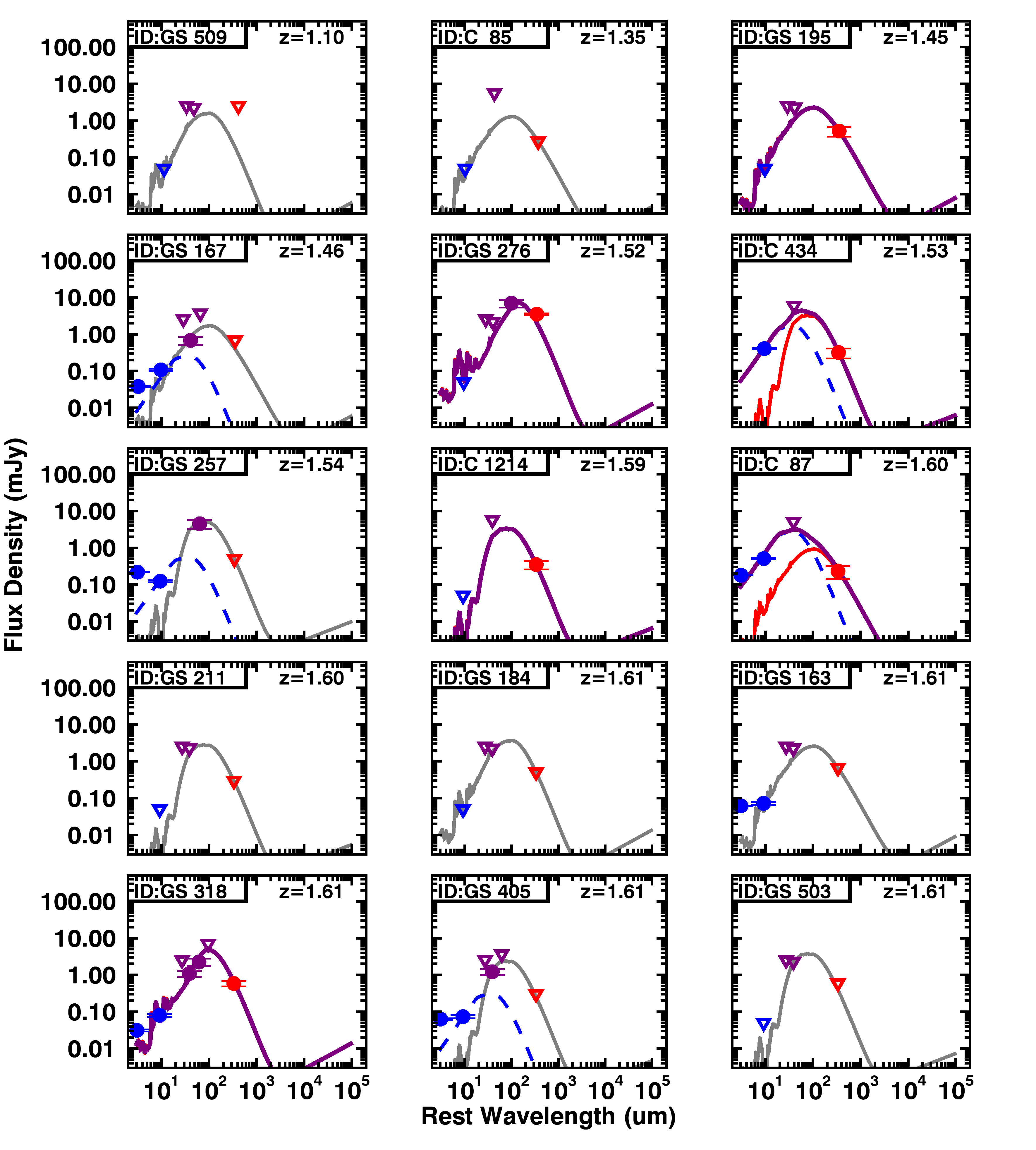}
		\caption{The best-fit SEDs for all sources in our X-ray AGN sample.
			Here we give the first 15 sources, the rest being available on the online version. 
		    The blue dashed curve is the AGN component, 
			while the red solid curve is the star-forming component. The total 
			combined SED is shown as a purple solid curve. The grey curves correspond 
			to an upper limit constraint on the SF component.
			The photometry is colour-coded, with blue corresponding to {\it Spitzer}, 
			purple to {\it Herschel} bands, and red to the ALMA photometry.
			Filled circles correspond to photometric measurements, while the
			inverted triangles correspond to photometric upper limits. We note that here we plot all AGN 
			components found in our SED fitting analysis, including weak/uncertain ones 
			(see flag = 1 in Tables~A1\&A2) that where not included in our analysis.}\label{bestfits}
	\end{center}
\end{figure*}

\setcounter{figure}{1}
\begin{figure*}  
	\begin{center}
		\includegraphics[scale=1.2]{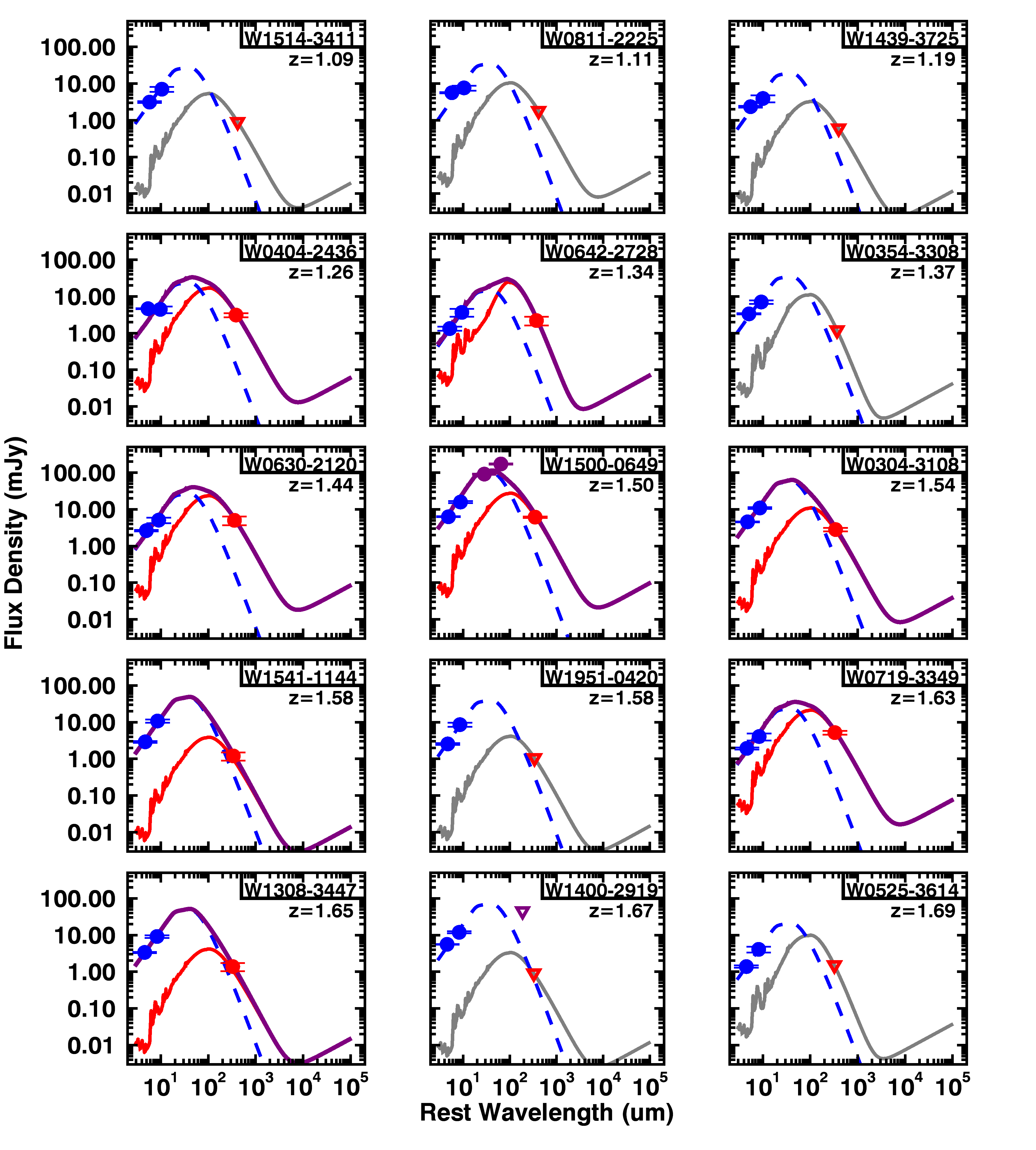}
		\caption{The best-fit SEDs for all sources of the MIR-bright AGN comparison sample 
		at redshifts of $1<z<5$. Here we give the first 15 sources, 
		the rest being available on the online version. 
		The blue dashed curve is the AGN component, 
			while the red solid curve is the star-forming component. The total 
			combined SED is shown as a purple solid curve. The grey curves correspond 
			to an upper limit constraint on the SF component.
			The photometry is colour-coded, with blue corresponding to {\it Spitzer}, 
			purple to {\it Herschel} bands, and red to the ALMA photometry. 
			Filled circles correspond to photometric measurements, while the
			 inverted triangles correspond to photometric upper limits. }\label{bestfits_lonsdale}
	\end{center}
\end{figure*}

\setcounter{figure}{2}
\begin{figure*}  
	\begin{center}
		\includegraphics[scale=1.2]{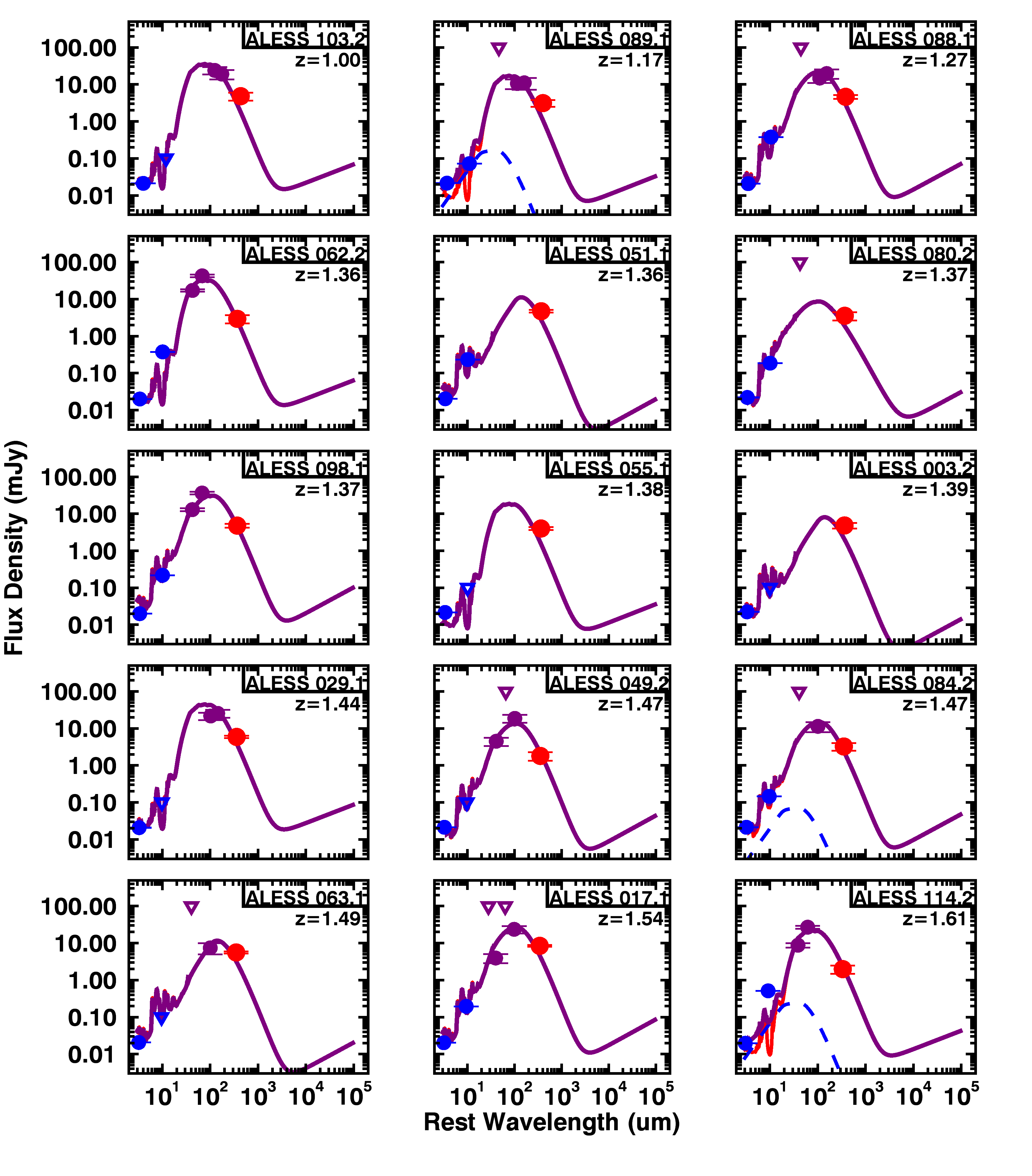}
		\caption{The best-fit SEDs for all sources of the ALESS SMG comparison sample at redshifts of 
		$1<z<5$. Here we give the first 15 sources, the rest being available on the online version. 
		The blue dashed curve is the AGN component, 
			while the red solid curve is the star-forming component. The total 
			combined SED is shown as a purple solid curve. The grey curves correspond 
			to an upper limit constraint on the SF component.
			The photometry is colour-coded, with blue corresponding to {\it Spitzer}, 
			purple to {\it Herschel} bands, and red to the ALMA photometry. 
			Filled circles correspond to photometric measurements, while the
			 inverted triangles correspond to photometric upper limits. We note that here we plot all AGN 
			components found in our SED fitting analysis, including weak/uncertain ones 
			(see flag = 1 in Table~A4) that where not included in our analysis.}\label{bestfits_aless}
	\end{center}
\end{figure*}

\end{document}